%% file: ms.tex
\documentclass[usenatbib]{mn2e}
\usepackage{graphicx,times, amsmath, epsfig}

\include{defns}

\pdfoutput=1

\begin{document}

\title[Detecting the peaks of the cosmological 21cm signal]{Reionization and Beyond: detecting the peaks of the cosmological 21cm signal}

\author[Mesinger et al.]{Andrei Mesinger$^1$\thanks{email: andrei.mesinger@sns.it}, Aaron Ewall-Wice$^2$, \& Jacqueline Hewitt$^2$\\
$^1$Scuola Normale Superiore, Piazza dei Cavalieri 7, 56126 Pisa, Italy\\
$^2$Massachusetts Institute of Technology, Cambridge, MA 02139, USA
}

\voffset-.6in

\maketitle

\begin{abstract}
The cosmological 21cm signal is set to become the most powerful probe of the early Universe, with first generation interferometers aiming to make statistical detections of reionization.  There is increasing interest also in the pre-reionization epoch when the intergalactic medium (IGM) was heated by an early X-ray background.
Here we perform parameter studies varying the halo masses capable of hosting galaxies, and their X-ray production efficiencies.   These two fundamental parameters control the timing and relative offset of reionization and IGM heating, making them the most relevant for predicting the signal during both epochs. We also relate these to popular models of Warm Dark Matter cosmologies. For each parameter combination we compute the signal-to-noise (S/N) of the large-scale ($k\sim0.1$Mpc$^{-1}$) 21cm power for both reionization and X-ray heating for a 2000h observation with several instruments: 128 tile Murchison Wide Field Array (MWA128T), a 256 tile extension (MWA256T), the Low Frequency Array (LOFAR), the 128 element Precision Array for Probing the Epoch of Reionization (PAPER), and the second generation Square Kilometre Array (SKA). We show that X-ray heating and reionization in many cases are of comparable detectability.  For fiducial astrophysical parameters, MWA128T might detect X-ray heating thanks to its extended bandpass. When it comes to reionization,  both MWA128T and PAPER will also only achieve marginal detections, unless foregrounds on larger scales can be mitigated.  On the other hand, LOFAR should detect plausible models of reionization at S/N $>10$. The SKA will easily detect both X-ray heating and reionization.
\end{abstract}

\begin{keywords}
cosmology: theory -- intergalactic medium -- early Universe -- reionization -- dark ages -- X-rays:galaxies
\end{keywords}

\section{Introduction}
\label{sec:intro}

The dawn of the first stars and black holes of our Universe is at the forefront of modern cosmological research.  The redshifted 21cm line from neutral hydrogen will arguably provide the largest insights into these epochs.  The 21cm signal is sensitive to the ionization and thermal state of the gas, and is therefore a powerful probe of both the intergalactic medium (IGM; where most of the baryons reside), as well as the first galaxies (whose radiation governs the evolution of the IGM).  Since it is a line transition, the 21cm signal can tell us about the three dimensional structure of cosmic gas, and its evolution.  First generation interferometers, like the Low Frequency Array (LOFAR; \citealt{vanHaarlem13})\footnote{http://www.lofar.org/},  Murchison Wide Field Array (MWA; \citealt{Tingay12})\footnote{http://www.mwatelescope.org/}, and the Precision Array for Probing the Epoch of Reionization (PAPER; \citealt{Parsons10})\footnote{http://eor.berkeley.edu} are coming on-line, with second-generation instruments such as the Square Kilometre Array (SKA; \citealt{SKA12})\footnote{http://www.skatelescope.org/} soon to follow, offering the promise of full tomographical imaging of the early Universe.

  Given that initial interferometric measurements will likely be noise-limited, the first-generation instruments are focusing on statistical detections: going after the large-scale, spherically-averaged 21cm power spectrum.
  Furthermore, efforts have mostly focused on the reionization epoch.
However, it is highly likely that the peak in the amplitude of large-scale fluctuations occurred during the preceding epoch when X-rays began heating the cold IGM (e.g. \citealt{PF07, MF07, Baek10, Santos10, MO12, MFS13}).  Sourced by strong absorption of cold gas against the CMB and large temperature fluctuations in the IGM, the 21cm power during X-ray heating is expected to be at least an order of magnitude higher than that during reionization.  As we shall see below, in many cases this increase is large enough to compensate for the increase in the thermal noise of the interferometer at the corresponding lower frequencies. Considering sensitivity alone, detecting the heating epoch in 21cm interferometry can therefore be of comparable difficulty to detecting reionization, though the potential challenges of radio frequency interference (RFI; the heating epoch extends through the FM band) and calibrating a larger beam might pose additional challenges. 

Heating is expected to be dominated by the X-rays from early astrophysical sources, most likely X-ray binaries (XRB; e.g. \citealt{Mirabel11, Fragos12}).  However, some classes of popular annihilating dark matter (DM) models can also imprint a strong signature in the IGM thermal evolution, which is not degenerate with that of the astrophysical X-rays (e.g. \citealt{Chuzhoy08, Valdes13}, Evoli et al., in prep)\footnote{ Initially, other sources of heating were thought to be important, sourced by the Ly$\alpha$ background \citep{MMR97} and structure formation shocks (e.g. \citealt{GS04}).  However, these are now thought to be sub-dominant to X-rays (e.g. \citealt{CM04, Rybicki06, FL04, MO12}).  In particular, \citet{MO12} perform convergence tests, using both grid and smoothed particle hydrodynamics simulations, quantifying the importance of shock heating in their Appendix A.  They find that shock heating only boosts the mean temperature by a few percent at $z\gsim12$, thus having a negligible impact on our conclusions below.  Extreme models in which heating and reionization occurs very late ($z<10$), could have a somewhat larger contribution from shock heating ($\sim$10\%), and we caution the reader not to over-interpret the precise values of the signal in this, admittedly unlikely, regime.}
Therefore the 21cm power spectrum during the heating regime encodes valuable astrophysical and even cosmological insight.

In this work, we quantify the detectability of {\it both} X-ray heating {\it and} reionization with upcoming and future 21cm interferometers.  We perform an astrophysical parameter study, exploring different minimum DM halo masses required to host galaxies, $\Mmin$, as well as the galactic X-ray emissivity. We also discuss the observability of the signal in terms of popular warm dark matter models, recasting $\Mmin$ to an analogous warm dark matter particle mass, $\Mwdm$. During the completion of this work, a similar study was presented by \citet{CL13}. Our work extends their results, including broader astrophysical parameter space exploration, furthering physical intuition, and incorporating sensitivity models of several upcoming interferometers.

We focus on detecting the power spectrum in a single $k$-bin, centered around $k=0.1$ Mpc$^{-1}$, which \citet{Pober13} identifies as being relatively clean of foregrounds.  Digging out larger scales will likely require extensive foreground cleaning, while smaller scales are quickly drowned by instrument noise (we find a factor of $\sim$5 increase in the rms noise going to $k=0.2$ Mpc$^{-1}$).  Hence detections with the first generation instruments might have only a relatively narrow window available in $k$ space.

This paper is organized as follows. In \S \ref{sec:cosmo} we describe our simulations of the cosmological signal, while in \S \ref{sec:sens} we discuss the adopted telescope sensitivities.  In \S \ref{sec:results}, we present our results, including the detectability of the peak signal and peak S/N across our parameter space. In \S \ref{sec:unsmooth} we briefly consider the potential impact of foreground contamination.  Finally we conclude in \S \ref{sec:conc}.

Unless stated otherwise, we quote all quantities in comoving units. We adopt the background cosmological parameters: ($\Omega_\Lambda$, $\Omega_{\rm M}$, $\Omega_b$, $n$, $\sigma_8$, $H_0$) = (0.68, 0.32, 0.049, 0.96, 0.83, 67 km s$^{-1}$ Mpc$^{-1}$), consistent with recent results from the Planck mission \citep{Planck13}.

\section{Cosmological signal}
\label{sec:cosmo}

The 21cm signal is usually represented in terms of the offset of the 21cm brightness temperature from the CMB temperature, $\Tcmb$, along a line of sight (LOS) at observed frequency $\nu$ (c.f. \citealt{FOB06}):
\begin{align}
\label{eq:delT}
\nonumber \delT(\nu) = &\frac{\Ts - \Tcmb}{1+z} (1 - e^{-\tau_{\nu_0}}) \approx \\
\nonumber &27 \nf (1+\delNL) \left(\frac{H}{dv_r/dr + H}\right) \left(1 - \frac{\Tcmb}{\Ts} \right) \\
&\times \left( \frac{1+z}{10} \frac{0.15}{\Omega_{\rm M} h^2}\right)^{1/2} \left( \frac{\Omega_b h^2}{0.023} \right) {\rm mK},
\end{align}
\noindent where $T_S$ is the gas spin temperature, $\tau_{\nu_0}$ is the optical depth at the 21-cm frequency $\nu_0$, $\delNL({\bf x}, z) \equiv \rho/\bar{\rho} - 1$ is the evolved (Eulerian) density contrast, $H(z)$ is the Hubble parameter, $dv_r/dr$ is the comoving gradient of the line of sight component of the comoving velocity, and all quantities are evaluated at redshift $z=\nu_0/\nu - 1$.

To simulate the 21cm signal, we use a parallelized version of the publicly available \cmfast\ code\footnote{http://homepage.sns.it/mesinger/Sim.html}. \cmfast\
 uses perturbation theory (PT) and excursion-set formalism to generate density, velocity, source, ionization, and spin temperature fields.  For further details and tests of the code, interested readers are encouraged to see \citet{MF07}, \citet{Zahn11}, \citet{MFC11} and \citet{MFS13}.  Here we outline our simulation set-up and the free parameters in our study.

Our simulation boxes are 600 Mpc on a side, with a resolution of 400$^3$.  Ionizations by UV photons are computed in an excursion-set fashion \citep{FZH04}, by comparing the local number of ionizing photons to neutral atoms.  The cumulative number of ionizing photons is given by multiplying the fraction of mass collapsed in halos more massive than some threshold mass,  $f_{\rm coll}(>\Mmin)$, by an ionizing efficiency which can be written as:
\begin{equation}
\label{eq:f_UV}
\zeta_{\rm UV} = 30  \bigg(\frac{N_\gamma}{4400}\bigg) \bigg(\frac{f_{\rm esc}}{0.1}\bigg) \bigg(\frac{f_\ast}{0.1}\bigg) \bigg(\frac{1.5}{1+\bar{n}_{\rm rec}}\bigg) ~ ,
\end{equation}
\noindent where $f_\ast$ is the fraction of gas converted into stars, $N_\gamma$ is the number of ionizing photons per stellar baryon, $f_{\rm esc}$ is the fraction of UV ionizing photons that escape into the IGM, and $\bar{n}_{\rm rec}$ is the mean number of recombinations per baryon.  Here we fix the ionizing efficiency to $\zeta_{\rm UV} = 30$, which agrees with the measured electron scattering optical depth in the fiducial model (defined below), and instead vary the X-ray luminosity of galaxies.  Although there is uncertainty in the value and evolution of $\zeta_{\rm UV}$, by varying $\Mmin$ we reasonably capture the redshift evolution of the reionization peak (corresponding to $\avenf\sim0.5$), which is the dominant factor in its detectability (along with the offset of the reionization and heating epochs).  Hence, even though our main focus here is the 21cm peak power which is very insensitive to changes in $\zeta_{\rm UV}$ (e.g. \citealt{MFS13, CL13}), we expect that our range of reionization S/N estimates to also be robust.

The comoving X-ray emissivity in our models can be expressed as
\begin{align}
\label{eq:emissivity}
\nonumber \epsilon_{h \nu}(\nu_e, {\bf x}, z) = &\alpha h \frac{N_{\rm X}}{\mu m_p} \left( \frac{\nu_e}{\nu_0} \right)^{-\alpha} \\
&\left[ \rho_{\rm crit, 0} \Omega_b f_\ast (1+\bar{\delta}_{\rm nl}) \frac{d f_{\rm coll}(>\Mmin)}{dt} \right] ~ ,
\end{align}
\noindent where $N_{\rm X}$ is the number of X-ray photons per stellar baryon, $\mu m_p$ is the mean baryon mass, $\rho_{\rm crit, 0}$ is the current critical density, $f_\ast$ is fraction of baryons converted into stars (we take $f_\ast=0.1$), $\bar{\delta}_{\rm nl}$ is the mean non-linear overdensity.  The quantity in the brackets is the comoving star formation rate density (SFRD).  We assume the same $\Mmin$ for both UV and X-ray sources. Following previous works (e.g. \citealt{PF07, Santos08, Baek10, MFC11}), we take a spectral (energy) index of $\alpha=1.5$, and assume photons below  $h \nu_0=$ 300 eV are obscured, these having optical depths exceeding unity for $N_{\rm HI}\gsim10^{21.5}$ cm$^{-2}$, consistent with the column densities seen in high-redshift gamma-ray bursts (GRBs; \citealt{Totani06, Greiner09}).  In addition to heating by X-rays, we include Compton heating, adiabatic cooling/heating, and heating through changing ionization species \citep{MFC11}.

We compute the Wouthuysen-Field (WF; \citealt{Wouthuysen52, Field58}) coupling (i.e. Ly$\alpha$ pumping; when the Ly$\alpha$ background from the first stars couples the spin temperature to the gas temperature) using the Lyman resonance backgrounds from both X-ray excitation of HI, and direct stellar emission.  The later is found to dominate by two orders of magnitude in our fiducial models.  For the direct stellar emission, we assume standard Population II spectra from \citep{BL05_WF}, and sum over the Lyman resonance backgrounds \citep{MFC11}. This fiducial  spectrum results in $\sim10^4$ rest-frame photons between \lya\ and the Lyman limit.

Our models have two free parameters:

\begin{enumerate}

\item {\bf $f_X$} -- {\it the X-ray efficiency of galaxies}.  Our fiducial choice of $f_X \equiv (N_{\rm X}/0.25) = 1$ corresponds to $N_{\rm X}=0.25$ X-ray photons per stellar baryon.  This choice results from a total X-ray luminosity above $h\nu_0=0.3$ keV of $L_{\rm X, {\rm 0.3+keV}}\sim10^{40}$ erg s$^{-1}$ ($\Msun$ yr$^{-1})^{-1}$, using our spectral energy index of $\alpha=1.5$.\footnote{It is more common in the literature to parameterize X-ray efficiency by the ratio of the X-ray luminosity to star formation rate, generally measured for star burst galaxies.  However, this number depends on the choice of bandwidth over which the X-ray luminosity is measured (e.g. \citealt{MGS12}).  We also note that many of the observationally quoted X-ray luminosities are sensitive only to energies high enough to interact little with the IGM (for example, photons with energies $\gsim2$ keV have mean free paths greater than the Hubble length at $z\sim15$, even through a neutral Universe; \citealt{Baek10, McQuinn12}).  Hence the value of $f_X$ for even low redshift galaxies is very uncertain.}
 This choice is consistent with (a factor of $\sim$2 higher than) an extrapolation from the 0.5--8 keV measurement of \citet{MGS12}, $L_{\rm X, {\rm 0.5-8keV}}\approx3\times10^{39}$ erg s$^{-1}$ ($\Msun$ yr$^{-1})^{-1}$.  It is highly uncertain how the X-ray luminosity evolves towards higher redshifts, although several studies argue that the higher binary fraction expected in the first galaxies results in more XRBs (e.g. \citealt{Mirabel11, Fragos12}).  There is also tentative evidence from the Chandra Deep Field-South (CDF-S; e.g. \citealt{Xue11}) that the X-ray luminosity to star formation rate is increasing out to  $z\sim4$ \citep{Basu-Zych13}.  Extreme evolution is limited by the 2$\sigma$ upper limits from the $z\sim6$ galaxy sample of \citet{CBH12}, which admittedly still allow $N_{\rm X}\sim1000$, e.g. assuming a SFR of $\sim$0.1 $\Msun$ yr$^{-1}$, and our fiducial choice of $h\nu_0=0.3$ keV and $\alpha=1.5$.  {\it Below, we explore the reasonable range $10^{-3} \lsim f_X \lsim 10^3$.}\\

\item {\bf $\Mmin$} -- {\it the minimum mass of DM halos which host star-forming galaxies.}  $\Mmin$ can be expressed as:
\begin{align}
\label{eq:Mmin}
\Mmin & = 10^{8}h^{-1} \left(\frac{\mu}{0.6}\right)^{-3/2} \left(\frac{\Omega_{\rm m}}{\Omega_{\rm m}^{\phantom{\text{ }}\rm z}} \frac{\Delta_{\rm c}}{18\pi^{2}}\right)^{-1/2} \times \nonumber \\ & \times \left(\frac{T_{\rm vir}}{1.98\times 10^{4}\text{ K}}\right)^{3/2} \left(\frac{1+z}{10}\right)^{-3/2} M_{\odot} \simeq \nonumber
\\  & \simeq 10^{8}\left(\frac{1+z}{10}\right)^{-3/2}M_{\odot}
\end{align}
\noindent where $\mu$ is the mean molecular weight, $\Omega_{\rm m}^{\phantom{\text{ }}\rm z}=\Omega_{\rm m}\left(1+z\right)^{3}/\left[\Omega_{\rm m}\left(1+z\right)^{3}+\Omega_{\Lambda}\right]$ and $\Delta_{\rm c}=18\pi^{2}+82d-39d^{2}$ with $d=\Omega_{\rm m}^{\phantom{\text{ }}\rm z}-1$.\footnote{Another common approach is to argue that efficient cooling of gas at a redshift-independent temperature sets the threshold for hosting star forming galaxies.  This motivates using a fixed halo virial temperature, $\Tvir$, as a fundamental parameter, effectively introducing a redshift dependence to $\Mmin$ according to eq. \ref{eq:Mmin}.  However, using a fixed $\Mmin$ facilitates a more straightforward mapping to a particle mass in popular warm dark matter cosmologies, as we shall see below.
 In any case, both a fixed $\Mmin$ or fixed $\Tvir$ are oversimplifications, since feedback physics, either by SNe (e.g. \citealt{SH03}) or the X-ray and UV backgrounds themselves (e.g. \citealt{RO04, KM05, MFS13}), will likely govern the redshift evolution of $\Mmin$.  We are mostly interested in the value of $\Mmin$ during the X-ray heating phase.  It is unlikely there is dramatic evolution of $\Mmin$ during this relatively rapid epoch (compare for example the fiducial model in \citealt{MFS13} which uses a fixed $\Tvir$, with the one in the top panels of Fig. \ref{fig:evolution}, which uses a fixed $\Mmin$).  We note however that during the very early stages, a fixed $\Mmin$ model shows a more rapid evolution impacting the depth of the mean absorption trough (bottom panel of Fig. \ref{fig:evolution}) compared with a fixed $\Tvir$ model.} As a fiducial choice, we take $\Mmin\sim10^8\Msun$, corresponding to the atomic cooling threshold at $z\sim10$.  The first galaxies were likely hosted by less massive, molecularly cooled halos, $\Mmin\sim10^{6-7}\Msun$ (e.g., \citealt{HTL96, ABN02, BCL02}).  However, star formation inside such small halos was likely inefficient (with a handful of stars per halo), and was eventually suppressed by the heating from X-rays themselves or other feedback processes \citep{HAR00,RGS01,MBH06, HB06}.  $\Mmin$ could also have been larger than the atomic cooling threshold due to feedback processes (e.g. \citealt{SH03, OGT08, PS09, SM13b, SM13a}).   It is unlikely that $\Mmin$ was larger than $\sim10^{10-11}$ since these values approximately latch onto the steeply-rising faint end of the observed galaxy luminosity functions at $z\sim6$--8 (e.g., \citealt{SFD11, FDO11}).  Furthermore, it would be difficult to complete reionization by $z\sim$5--6 without a contribution from galaxies hosted by smaller halos (e.g. \citealt{KF-G12, CFG08}). {\it Below, we explore the reasonable range $10^7 \lsim \Mmin/\Msun \lsim 10^{10}$.}\\

\end{enumerate}

For the purposes of this work, it is useful to keep in mind that increasing $f_X$ has the effect of shifting the X-ray heating epoch (and associated peak in power) towards higher redshifts, while increasing $\Mmin$ has the effect of shifting {\it all} astrophysical epochs towards lower redshifts.  Furthermore, increasing $\Mmin$ has the additional impact of speeding up cosmic evolution, as structures form more rapidly on the high-mass end of the mass function.

\subsection{Warm Dark Matter Models}
\label{sec:WDM}

Our framework also allows us to estimate the 21cm signal in warm dark matter (WDM) cosmologies.   WDM models with particle masses of order $\Mwdm\sim$keV became popular as a cosmological way of alleviating small-scale problems of CDM, such as a dearth of locally observed dwarf galaxies and flattened rotation curves in galaxy centers (e.g. \citealt{Moore99}).  Current measurements place limits of $\Mwdm\gsim$1--3 keV \citep{BHO01, deSouza13, KMD13, PMH13, Viel13}, with various degrees of astrophysical degeneracy.  Due to the hierarchal nature of structure formation, the impact of WDM (or any model with a dearth of small-scale power) is larger at higher redshifts, with the Universe becoming increasingly empty.  Therefore, a detection of the pre-reionization 21cm signal could strengthen limits on $\Mwdm$ \citep{Sitwell14}.

Structure formation in WDM models is suppressed through (i) particle free-streaming, and (ii) residual velocity dispersion of the particles.  Effect (i) can be included by suppressing the standard matter transfer function below the free-streaming scale (e.g. \citealt{BOT01}), while effect (ii) acts as an effective pressure, slowing the early growth of perturbations (e.g. \citealt{BHO01}).  Effect (ii) is generally ignored as it is difficult to include in $N$-body simulations, since it translates to an {\it intra}-particle dispersion in the codes.  By an analogy to a baryonic Jeans mass, \citet{BHO01} derived a critical WDM halo mass for thermal relics, below which structure formation is suppressed due to the particle velocity dispersion.  \citet{deSouza13} empirically found (see their Fig. 1) that a step-function suppression of halos smaller than $\sim60$ times this critical Jeans mass, results in collapse fractions which are very close to the full random walk procedure of \citet{BHO01}.    A step-function suppression allows us to relate the WDM particle mass, $\Mwdm$ to an ``effective'' $\Mmin$:
\begin{equation}
M_{\rm min, eff} \approx 2\times10^{10} \Msun \left(\frac{\Omega_{\rm wdm}h^2}{0.15}\right)^{1/2}\left(\frac{m_{\rm wdm}}{1 \rm keV}\right)^{-4} \left(\frac{1+z_{\rm I}}{3500}\right)^{3/2} ~ , 
\label{eq:MWDM}
\end{equation}
\noindent where $z_{I}$ corresponds to the redshift of matter-radiation equality. With this casting we can present our results in terms of $\Mwdm$ as well as $\Mmin$, without running additional dedicated simulations.  There is however an important caveat: our simulations use a standard CDM transfer function \citep{EH98}, without the WDM cut-off.  The empirical calibration in eq. \ref{eq:MWDM} included the proper WDM transfer function \citep{BOT01}.  However, due to the sharpness of the barrier transition and strength of the effective pressure effect, it is not unreasonable to assume that most of the small-mass suppression is included, even with the CDM transfer function. We check the accuracy of this mapping below (see Fig. \ref{fig:evolution}) for fiducial astrophysical parameters.  We find that our simple prescription which underestimates the suppression by using the CDM power spectrum shifts the evolution of the 21cm signal to higher redshifts by a modest $\Delta z \approx 1$, with the peak power relatively unaffected (compare the $\Mwdm=2$keV magenta curve, with the $\Mmin = M_{\rm min, eff}=10^9\Msun$ red curve; the former includes the WDM transfer function of \citealt{BOT01}).  Therefore, we find this simple mapping of eq.(\ref{eq:MWDM}) reasonable, and include the corresponding $\Mwdm$ values on the right vertical axes of our plots.  However, we caution that this conversion should only be treated as approximate.

Furthermore, we stress that it is the {\it maximum} value of $\Mmin$, set by {\it either} cosmology {\it or} cooling physics, which regulates galaxy formation.   Hence if the gas cannot efficiently cool to form stars in halos down to the $M_{\rm min, eff}$ threshold from eq. (\ref{eq:MWDM}), we cannot use this signal to probe the WDM particle mass.

\section{Instrument sensitivity}
\label{sec:sens}

\begin{figure}
\vspace{-1\baselineskip}
{
\includegraphics[width=0.5\textwidth]{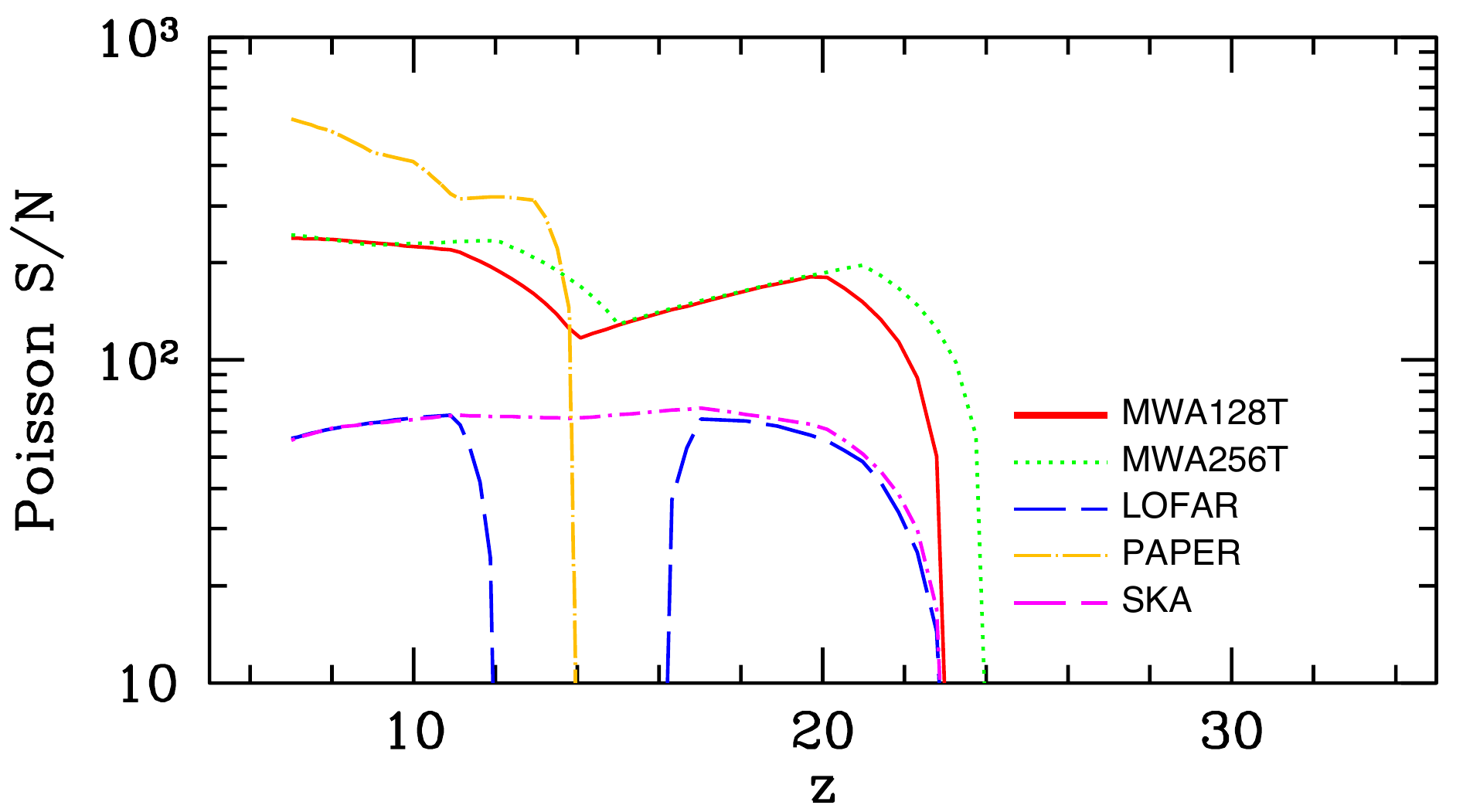}
}
\caption{
Poisson (cosmic variance) component of the S/N, i.e. $\sqrt N_k$, for our fiducial observational strategy and $k\approx0.1$ Mpc$^{-1}$.
\label{fig:poisson}
}
\vspace{-1\baselineskip}
\end{figure}

Throughout this work, we use as the observable quantity the spherically averaged power spectrum, $P_{21} \equiv k^3/(2\pi^2 V) ~ \bar{\delT}(z)^2 ~ \langle|\delta_{\rm 21}({\bf k}, z)|^2\rangle_k$, where $\delta_{21}({\bf x}, z) \equiv \delT({\bf x}, z)/ \bar{\delT}(z) - 1$.  Furthermore, we focus on the large-scale signal at $k=0.1$ Mpc$^{-1}$, lying in the ``sweet spot'' of 21cm interferometry: large enough for the cosmic signal not to be removed in the foreground cleaning process, yet small enough to have high S/N with upcoming instruments (e.g. \citealt{Lidz08, Dillon13, Pober13}).  Our default power spectrum bin width is $d \ln k =$ 0.5.

\subsection{Calculation of Thermal Noise}

To compute the thermal noise variance for each array, we perform rotation synthesis of 6 hours per night about zenith on a uv-plane whose resolution is set by the inverse of the array's primary beam FWHM. With the flat sky approximation, and assuming small baselines and  bandwidth, the amplitude of the noise power spectrum in each uv-cell is given by (e.g. \citealt{Morales05, Parsons12}):
\begin{equation}
P_N \approx D_M(z)^2Y\frac{k^3}{2 \pi^2} \frac{\Omega'}{2 t} T_{sys}^2,
\end{equation}
where $\Omega'$ is a beam dependent factor described in \citet{Parsons13} , $D_M(z)$ is a constant factor that converts between a transverse angle on the sky and comoving distance units, $Y$ is a factor that converts between frequency and radial comoving distance, and $t$ is the total time spent by all baselines in the uv cell during the aperture synthesis. $T_{sys}$ is the system temperature which is the sum of $T_{rec}$, the receiver noise temperature and $T_{sky}$, the sky temperature. For $T_{sky}$ we use the measurement of \citet{Rogers08} of $T_{sky} = 237 \left( \frac{\nu}{150 \text{MHz}} \right)^{-2.5} \text{K}$\footnote{We choose the minimal value measured at the galactic pole, where presumably power spectrum studies would focus. Measurements at hotter galactic latitudes could reduce our S/N estimates by a factor of 2-4.}. We set the receiver temperature to $T_{rec}=50 K + 0.1 T_{sky}$ \citep{Dewdney13}\footnote{While receiver temperature can vary from instrument to instrument, it is likely dominated by $T_{sky}$.}.

Chromatic effects due to the dependence of an array's uv coverage with frequency are approximated by averaging the coverage over the data cube and using the $T_{sys}$ at the data cube's center frequency. 
While mean noise power can be removed from the data by computing power spectra estimates, $\hat{P}_k$, from data interleaved in time \citep{Dillon13}, the variance of the thermal noise power spectrum is expected to be the leading contribution to measurement uncertainty within the ``EoR window". We assume that a power spectrum estimate is computed by taking an inverse variance weighted average of all u-v cells within a k-bin. In performing inverse variance weighting, the variance of a $\hat{P}_k$ is $\sigma_N^2 =1/\sum_j \sigma_j^{-2}$.

To account for foregrounds, we exclude all u-v cells lying within the ``wedge": the region of k-space contaminated by foregrounds which are thrown to larger $k_\parallel$ modes by the chromaticity of the interferometer. The maximum $k_\parallel$ contaminated by this mechanism, at a fixed $k_\perp$ is given by \citep{Morales12, Parsons12}:
\begin{equation}
k_\parallel^{max} = \sin \Theta_{max} \left( \frac{D_M(z) E(z)}{D_H (1+z)}\right) k_\perp + k_{intr} 
\end{equation}
where $D_H$ is the Hubble distance, $E(z) = H(z)/H_0$, and $\Theta_{max}$ is the maximum angle on the sky from which foregrounds enter the beam. $k_{intr}$ is an offset to account for the intrinsic ``spectral unsmoothness" of the foregrounds.
We take $k_{intr}=0.02$ Mpc$^{-1}$ and $\Theta_{max}$ equal to one half the FWHM of the primary beam. While this allows for possibly larger S/N at even larger scales ($k \sim 0.04$ Mpc$^{-1}$; e.g. \citealt{Beardsley13}), it is very uncertain whether this region will be clean of foregrounds. Hence we choose to compromise, working at what is more likely to be a foreground free scale, $k=0.1$ Mpc$^{-1}$ (e.g. \citealt{Pober13}). In \S \ref{sec:unsmooth} we examine sensitivity at smaller spatial scales with an even larger $k_{intr}$.


%

Although  8 MHz bands are a common choice in the literature, this bandwidth is large enough to average over signal evolution at very high redshifts.
 We therefore consider a band corresponding to $\Delta z = 0.5$ for all redshifts. In fixing the redshift span, we impose a minimal $k_\parallel$ resolution sampled by the instrument which becomes dramatically worse at high $z$. At $z \gsim 20$, this minimal $k_\parallel$ becomes so large that our instruments do not sample any modes at $k=0.1$Mpc$^{-1}$ (the sharp drop in the number of modes  at $z>20$ is evident in Fig. \ref{fig:poisson}).\footnote{Although necessary for sensitivity estimates, a bandwidth choice is relatively arbitrary.  Hence we do not additionally smooth the cosmic signal, showing its intrinsic value.  As already mentioned, our fiducial choice is motivated by the negligible evolution of the signal over the bandwidth, with the X-ray heating power near the peak evolving by only a few percent.  Nevertheless, we highlight that a wider bandwidth could extend SKA sensitivities to $z>20$.} 

We also include the Poisson noise of the cosmic signal, $P_{21}/\sqrt N_k$, where $N_k$ is the instrument-dependent number of modes in our power spectrum bin.  This Poisson term, corresponding to the maximum achievable S/N (i.e. if thermal noise is zero), is plotted in Fig. \ref{fig:poisson}.   Due to the large values of $N_k \sim10^3-10^4$, the Poisson (cosmic variance) noise only dominates high-sigma detections, generally achievable only with the SKA.  We note that our observational strategy is chosen to minimize the thermal noise.  A different strategy (sampling more independent fields) would lower the Poisson noise and increase the thermal noise; such a strategy can be employed to avoid the cosmic variance limited regime of SKA, if extremely high S/N is desired.
 For simplicity, we do not perform inverse weighing on the Poisson noise, noting that this has a minor impact (on very high S/N values).  Therefore, adding the noise terms in quadrature, our total S/N can be expressed as:
\begin{equation}
\label{eq:SN}
{\rm S/N} = \frac{P_{21}}{\sqrt{  P_{21}^2/N_k + \sigma_N^2}} ~ .
\end{equation}

\subsection{Array Models}

\begin{figure*}
\vspace{-1\baselineskip}
{
\includegraphics[width=\textwidth]{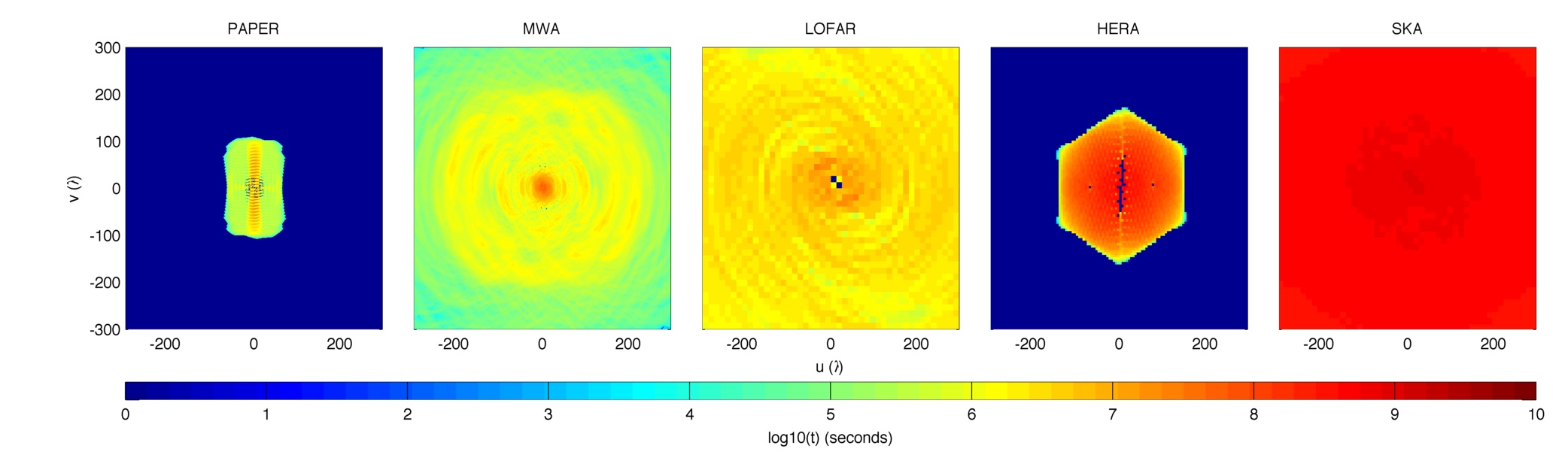}
}
\caption{
Time spent in each uv cell for a 2000h observation at $z=10$ for PAPER, MWA128T, LOFAR, HERA, SKA, ({\it left to right}).
\label{fig:UVcoverage}
}
\vspace{-0.5\baselineskip}
\end{figure*}

In this work, we consider four existent and planned arrays: MWA \citep{Tingay12}, LOFAR \citep{vanHaarlem13}, PAPER \citep{Parsons10}, and the SKA \citep{Dewdney13}. Here we briefly describe the model for each instrument.
\begin{description}
\item{{\bf MWA128T/256T}:} Our model of the MWA contains 128 tiles whose locations are given in \citet{Beardsley12}. Each antenna element is modeled by a 4x4 grid of short dipoles $0.3$ m above an infinite conducting plane and spaced 1.1 m apart. The primary beam FWHM is computed from the radiation pattern of this arrangement. The bandpass of the MWA cuts off at $\approx 75$ MHz so we set $P_N \to \infty$ below this frequency. A frequency resolution of 40 kHz is used.  We also consider a possible extension to 256 tiles (MWA256T), for which the infrastructure is already in place.  The additional tiles are placed randomly (with a 5 meter minimum separation), drawing from a uniform distribution within 50 meters of the center and then 1/$r^2$ for $r>50$ meters. 
\item{{\bf PAPER}:} We use the maximally redundant configuration of PAPER 128 with a band pass ranging from $100$ to $200$ MHz and a frequency resolution of 48 kHz. The primary beam is fixed to be 0.72 sr over all redshifts along with $\Omega'=1.69 sr$ \citep{Parsons13}. PAPER is a drift scan instrument, hence it observes a larger number of fields with a shorter integration time per field. We assume that 2000 hours of observation (an optimistically high choice corresponding to $\sim2$--3 observing seasons) are spread over observations of 3 different subfields with 2 hours of integration per field per night, increasing the Poisson counts significantly (Fig. \ref{fig:poisson}).
\item{{\bf LOFAR}:} LOFAR is comprised of two different sub-arrays. We use the high and low band antenna locations in the 40 core stations  along with the beams and effective areas described in \citet{vanHaarlem13}. For the low band antennas we assume the ``inner configuration", interpolating values for the beam FWHM and effective areas between those given in Table B.1 in \citet{vanHaarlem13}. We treat the region between the high and low bandpasses (80-110 MHz) as unobservable. A frequency resolution of 10 kHz is used.
\item{{\bf SKA}:} Antenna locations for our SKA model are based on the SKA Low Phase 1 design described in \citet{Dewdney13}. 866 station locations are drawn from a Gaussian distribution with 75 \% falling within 1000m of the center. Each station is modeled as a 17x17 array of log-periodic dipoles whose effective areas and beams are given in Table 3 and Appendix A of \citet{Dewdney13}. The frequency resolution is 1 kHz. 
\end{description}

In addition to the above fiducial instruments, we briefly present preliminary noise estimates from the proposed, second-generation Hydrogen Epoch of Reionization Arrays (HERA; http://reionization.org/).  HERA is a proposed instrument comprised of 547 antennas with a hexagonal packing configuration (Pober et al., in prep). Each of the static dish antennae is modeled as a 14 meter filled aperture. Because HERA would be a drift scan instrument with a narrow field of view, the observation strategy we adopt is to observe, on each night, 9 different $\sim$10 degree fields for 45 minutes each for a total of 6 hours per night. We adopt a frequency resolution of 98 kHz.  The configuration is optimized for statistical detections.

It is illustrative to compare the uv coverage of each array, especially within the compact core sourcing the sensitivity for the cosmological signal. We show in Figure \ref{fig:UVcoverage}, the time spent in each uv cell at $z=10$ for PAPER, MWA128T, LOFAR, HERA, SKA, ({\it left to right}), over the course of 2000 hours of observing. Similar plots for each individual array can be found in \citet{Beardsley13} (MWA), \citet{vanHaarlem13} (LOFAR) and \citet{Dewdney13} (SKA). PAPER and HERA, which are optimized for EoR measurements show compact uv distributions, while multi-purpose instruments like MWA, LOFAR, and the SKA have broad uv coverage. The MWA has a very wide field of view, resulting in fine pixelization. LOFAR, HERA, and SKA have relatively narrow primary beams, translating into a more coarse resolution in the uv plane.

\begin{figure*}
\vspace{-1\baselineskip}
{
\includegraphics[width=0.45\textwidth]{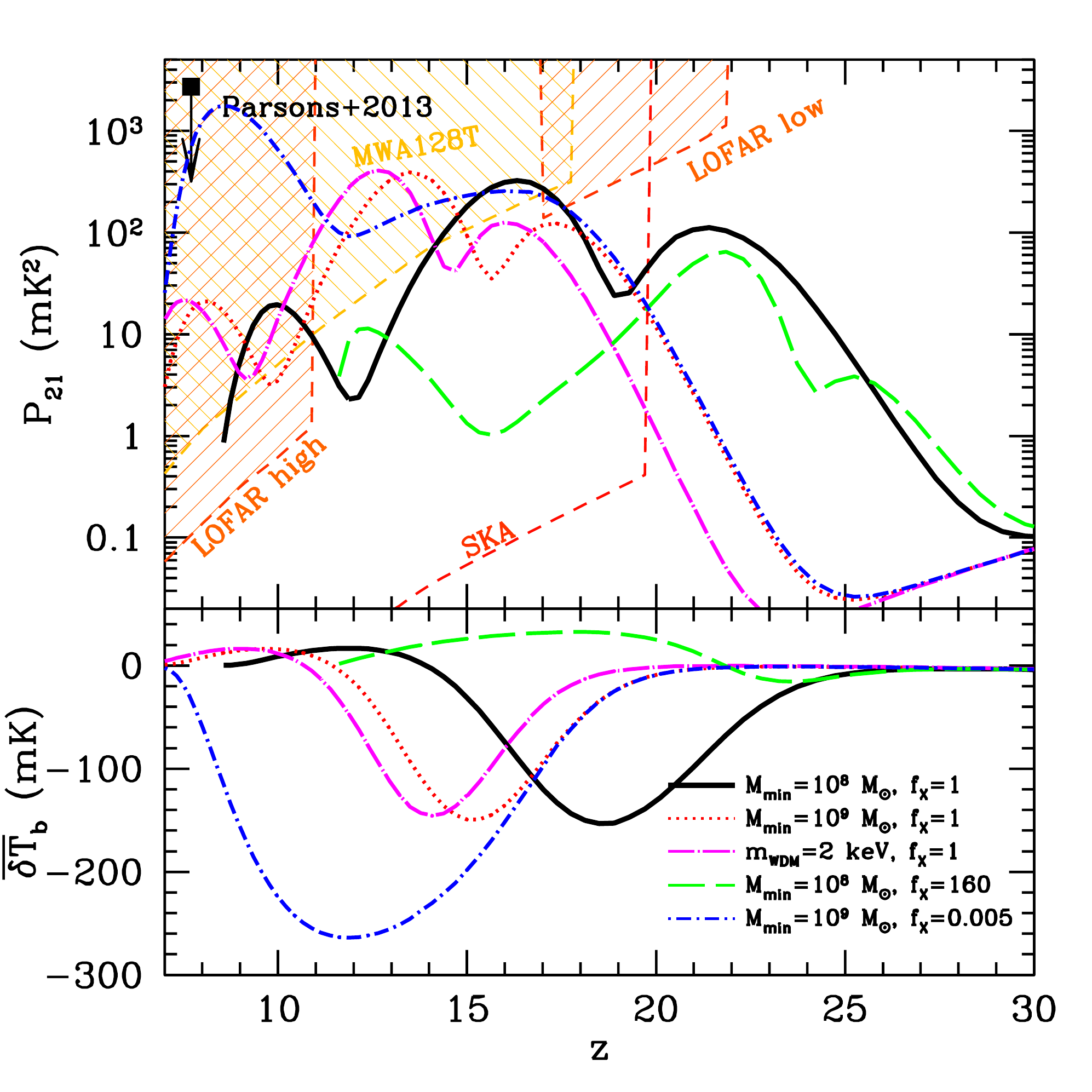}
\includegraphics[width=0.45\textwidth]{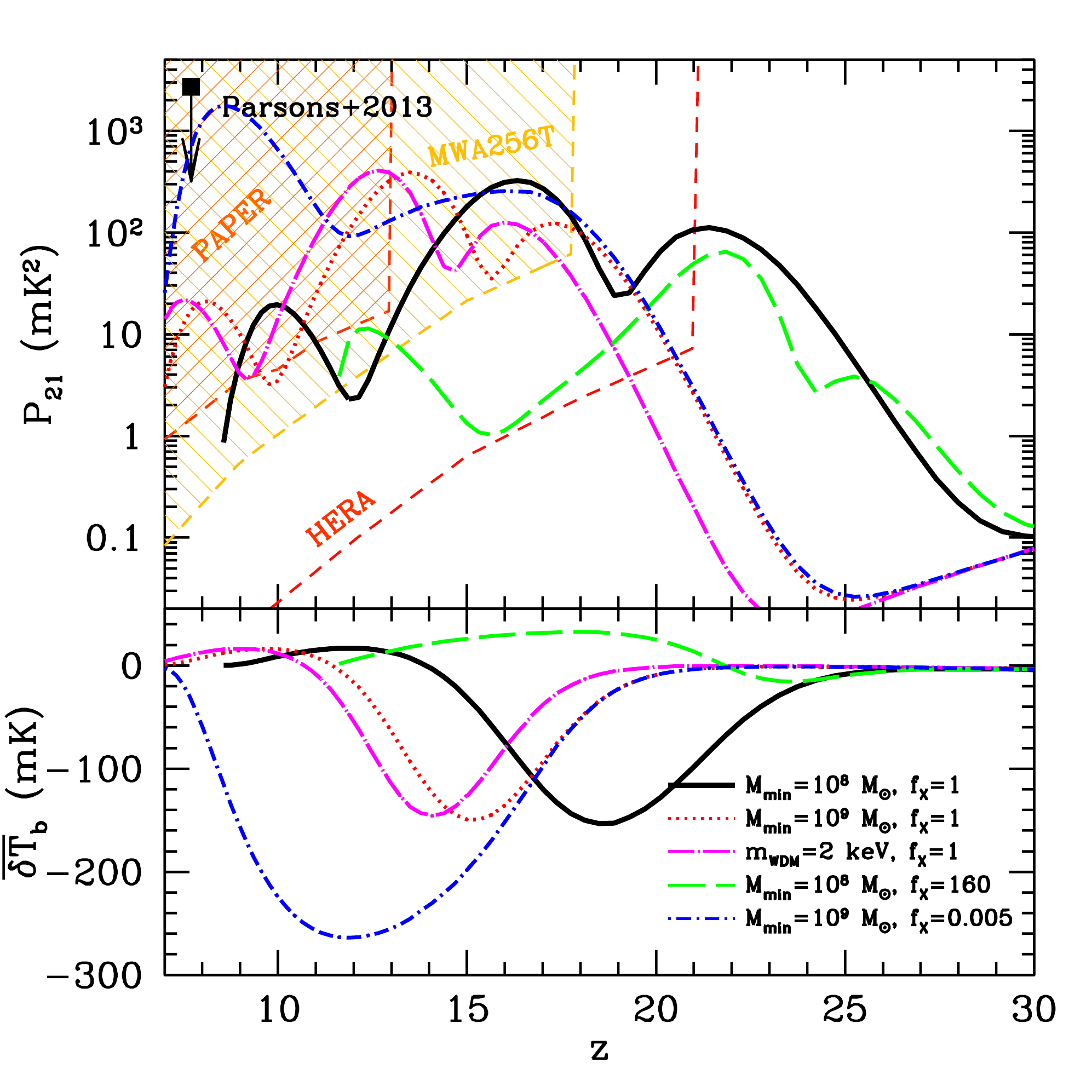}
}
\caption{
{\it Top panels:} Amplitude of the 21cm power at $k = 0.1$ Mpc$^{-1}$ in various models.  We also plot the (1$\sigma$) sensitivity curves corresponding to a 2000h observation with: MWA128T, LOFAR, SKA on the left, and MWA256T, PAPER, and the proposed HERA instrument on the right.  The recent upper limit from \citet{Parsons13} is shown at $z=7.7$.
 {\it Bottom panels:} The corresponding average 21cm brightness temperature offset from the CMB.
}
\label{fig:evolution}
\vspace{-1\baselineskip}
\end{figure*}

\section{Results}
\label{sec:results}

\subsection{Physical insight into the signal}

In the top panels of Fig. \ref{fig:evolution}, we plot the evolution of the $k = 0.1$ Mpc$^{-1}$ 21cm power in various models.  The black solid curves corresponds to a ``fiducial model'', with $f_X=1$ and $\Mmin=10^8 \Msun$ (approximately the atomic cooling threshold at these redshifts).

Note that most models exhibit the familiar three peak structure (e.g. \citealt{PF07, MFC11, Baek10}).  In order of decreasing redshift, these peaks correspond to the three astrophysical (radiation-driven) epochs of the 21cm signal: (i) WF coupling; (ii) X-ray heating; and (iii) reionization. 

Extreme models can avoid having a three-peaked structure by merging the reionization and X-ray heating peaks.  In these cases, the X-ray background is faint enough ($f_X \lsim 10^{-2}$; \citealt{MO12, CL13}), that it is unable to heat the IGM prior to the completion of reionization.  The resulting contrast between cold neutral and ionized regions can drive up the reionization peak considerably.  We show one such model with the blue curves in Fig. \ref{fig:evolution}.

We also overlay the sensitivity curves of MWA128T\footnote{Our MWA128T noise estimates are consistent with those in \citep{Beardsley13} at $z=8$, when accounting for their different choice of system temperature.  However, our noise curves are a factor of $\sim$10 higher than the ones in \citet{CL13}.  This is due primarily to the fact that they neglect to evolve the system temperature, using the $z=8$ value at all redshifts.  As the system temperature is expected to scale as $\propto (1+z)^{2.5}$, neglecting its evolution results in a dramatic underestimate of the noise at high-$z$}, LOFAR and SKA ({\it left panel}), as well as MWA256T, PAPER and the proposed HERA instrument
 ({\it right panel}).  The first generation instruments will have difficulty detecting the X-ray heating peak in the fiducial model. As the X-ray efficiency is decreased, or galaxies are hosted by more massive later-appearing halos, the heating peak moves to lower redshifts, making it more easily observable with the MWA (LOFAR unfortunately has a band gap in this regime, and the PAPER bandpass cuts-off beyond $z\approx13$).
 The SKA, within its bandpass, easily has the required sensitivity to detect all reasonable models, and is in fact limited by cosmic variance for our observational strategy, as we shall see below.\footnote{We caution that the effective bandpass for first generation instruments is limited by data transfer and processing, and will be narrower than the full range shown in Fig. \ref{fig:evolution}.  For example, MWA has an effective bandpass of 32MHz, allowing simultaneous observations over, e.g. $z=$8--10 and $z=$12--17.  Below, for simplicity and in order to be conservative, we only present S/N estimates in a single $\Delta z=0.5$ frequency bin. With wider frequency coverage, the S/N could be increased by a factor of a few by averaging over the effective bandpass.}

\begin{figure*}
{
\includegraphics[width=0.45\textwidth]{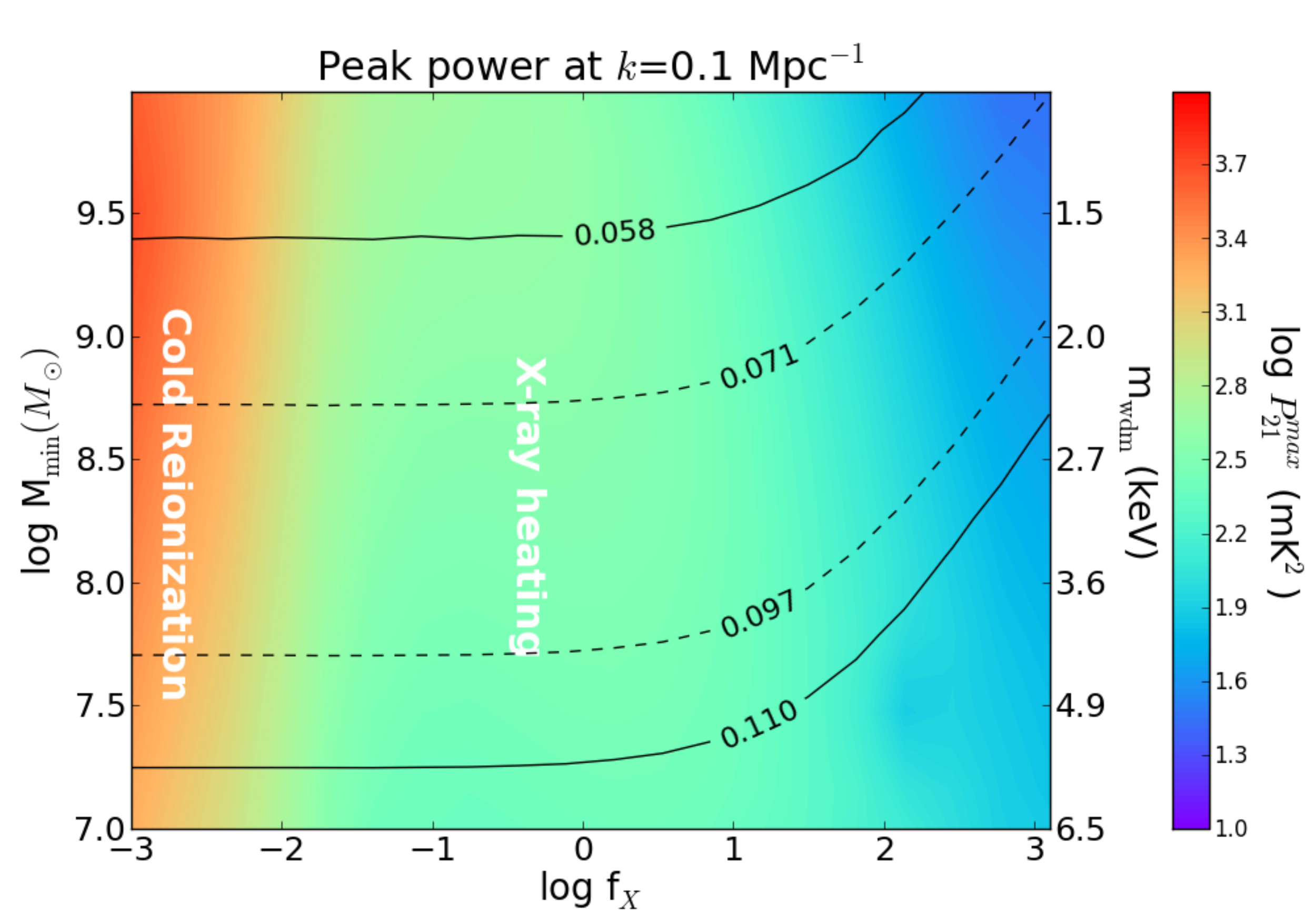}
\includegraphics[width=0.45\textwidth]{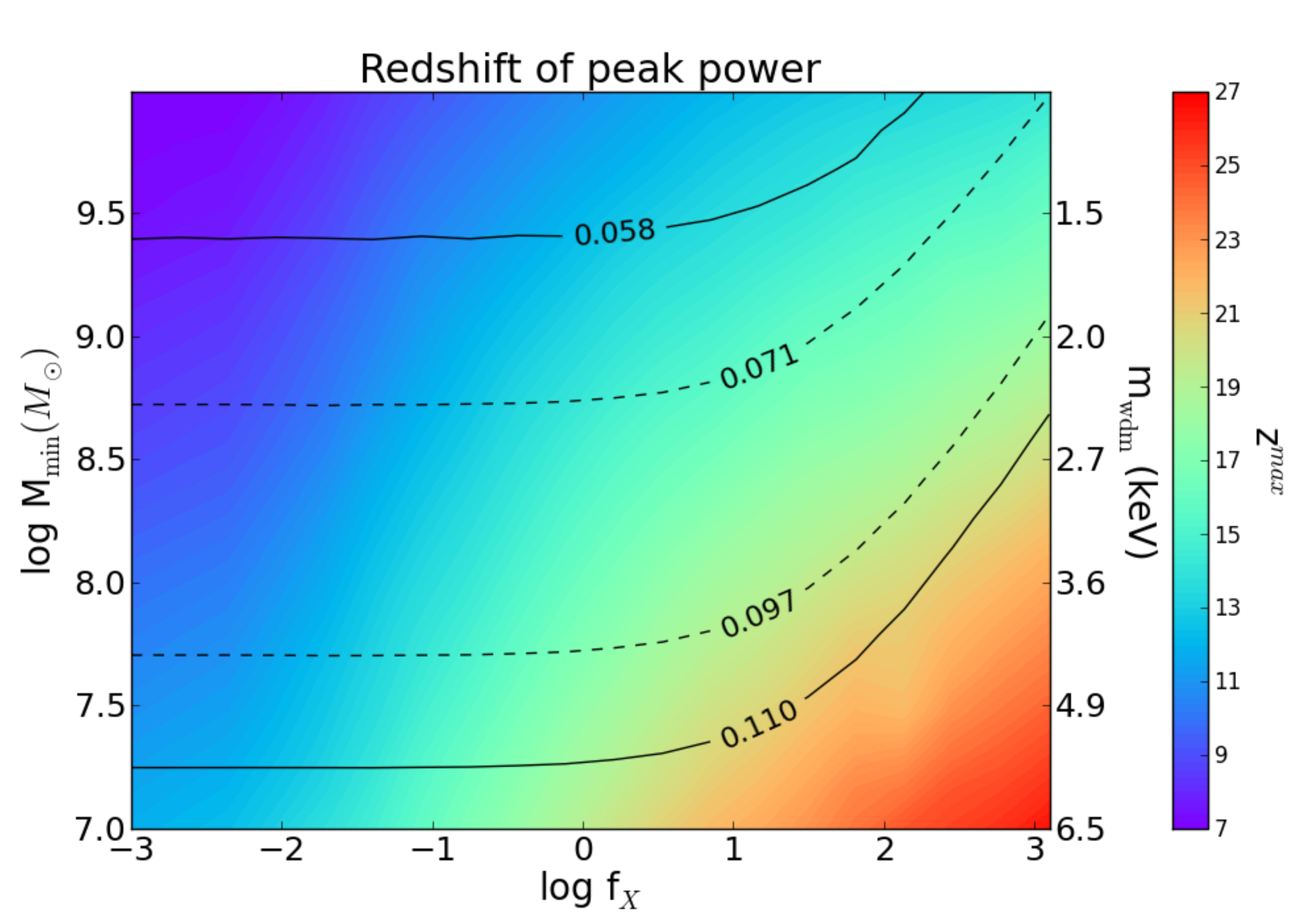}
\includegraphics[width=0.45\textwidth]{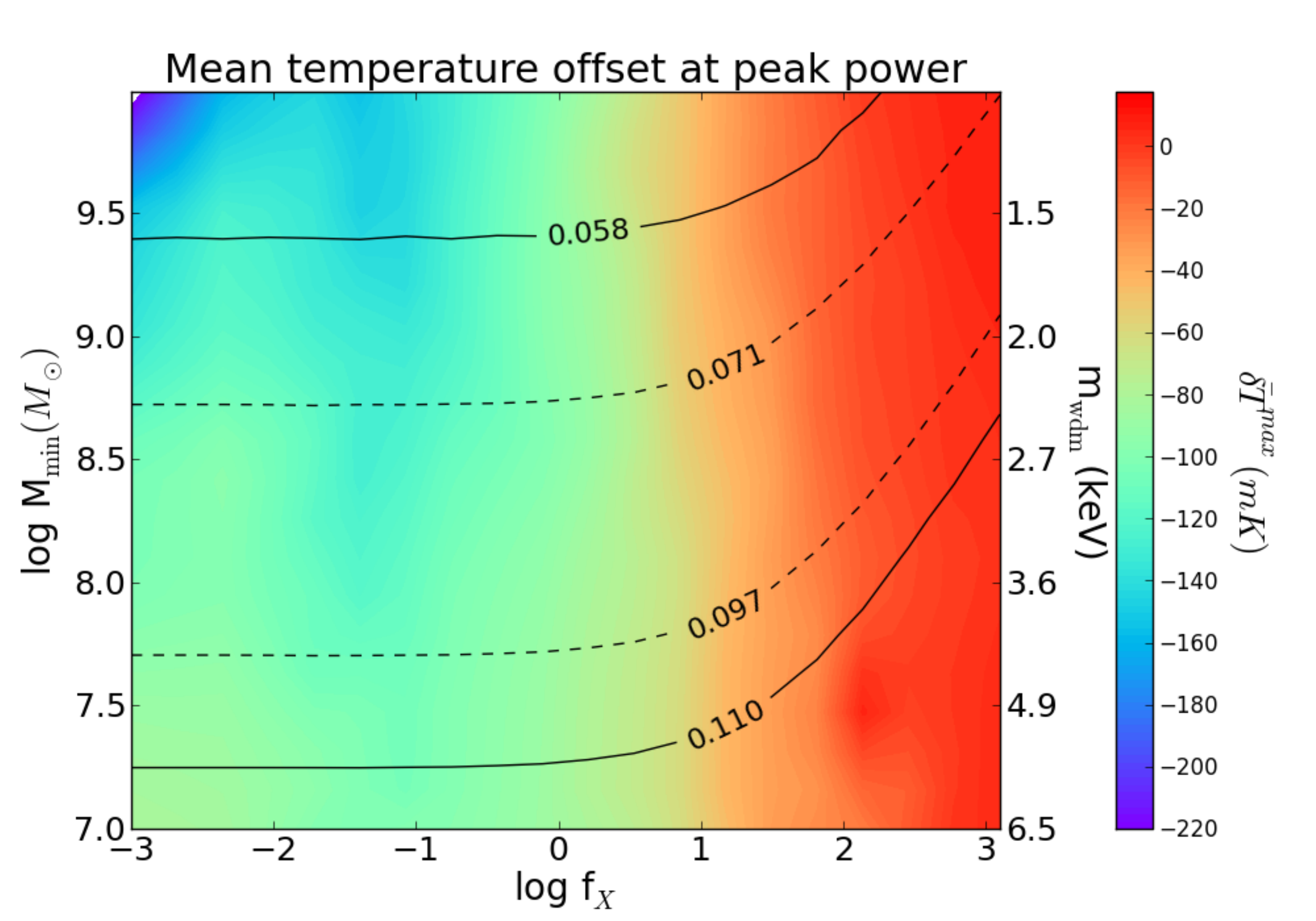}
\includegraphics[width=0.45\textwidth]{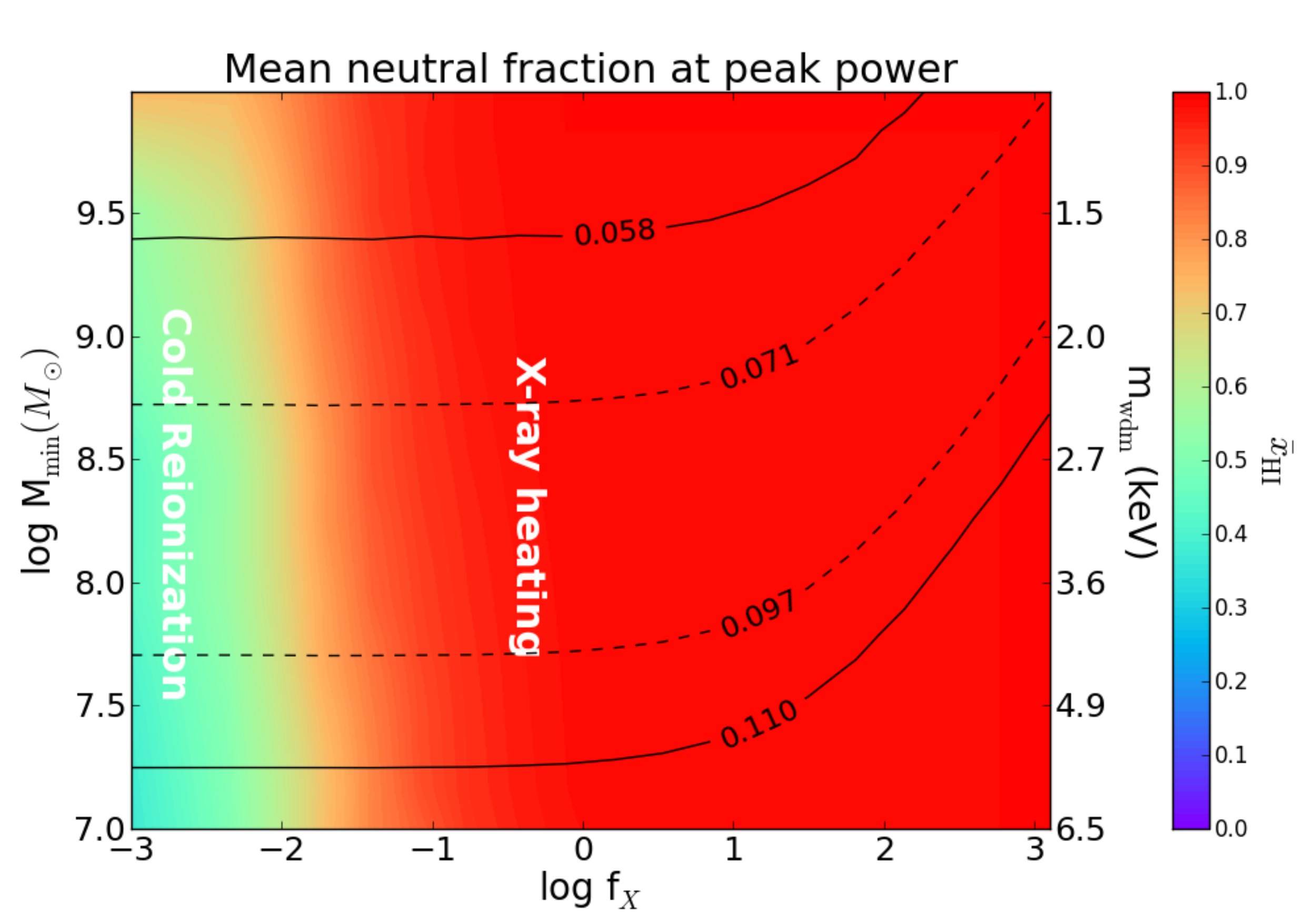}
}
\caption{
Various quantities evaluated at the redshift where the $k=0.1$ Mpc$^{-1}$ power is the largest in each model.  Shown are power amplitude, redshift, mean neutral fraction, and mean brightness temperature, clockwise from top left.  Overlaid are the Thompson scattering optical depth, $\tau_e$,
 contours corresponding to 1 and 2 $\sigma$ constraints from the 9yr release of WMAP data (WMAP9; \citealt{Hinshaw13}).  The right-side y-axis shows the corresponding values of the WDM particle mass, $\Mwdm$, computed according to the approximation in eq. (\ref{eq:MWDM}).  Regions of the parameter space in which the peak power occurs during the reionization or X-ray heating epochs are demarcated with the appropriate labels in some panels.
\label{fig:max_values}
}
\vspace{-1\baselineskip}
\end{figure*}

In Fig. \ref{fig:max_values}, we plot the power amplitude, redshift, mean neutral fraction, and mean brightness temperature (clockwise from top left), all evaluated at the redshift where the $k=0.1$ Mpc$^{-1}$ power is the largest in each model.

\begin{figure}
\vspace{-1\baselineskip}
{
\includegraphics[width=0.5\textwidth]{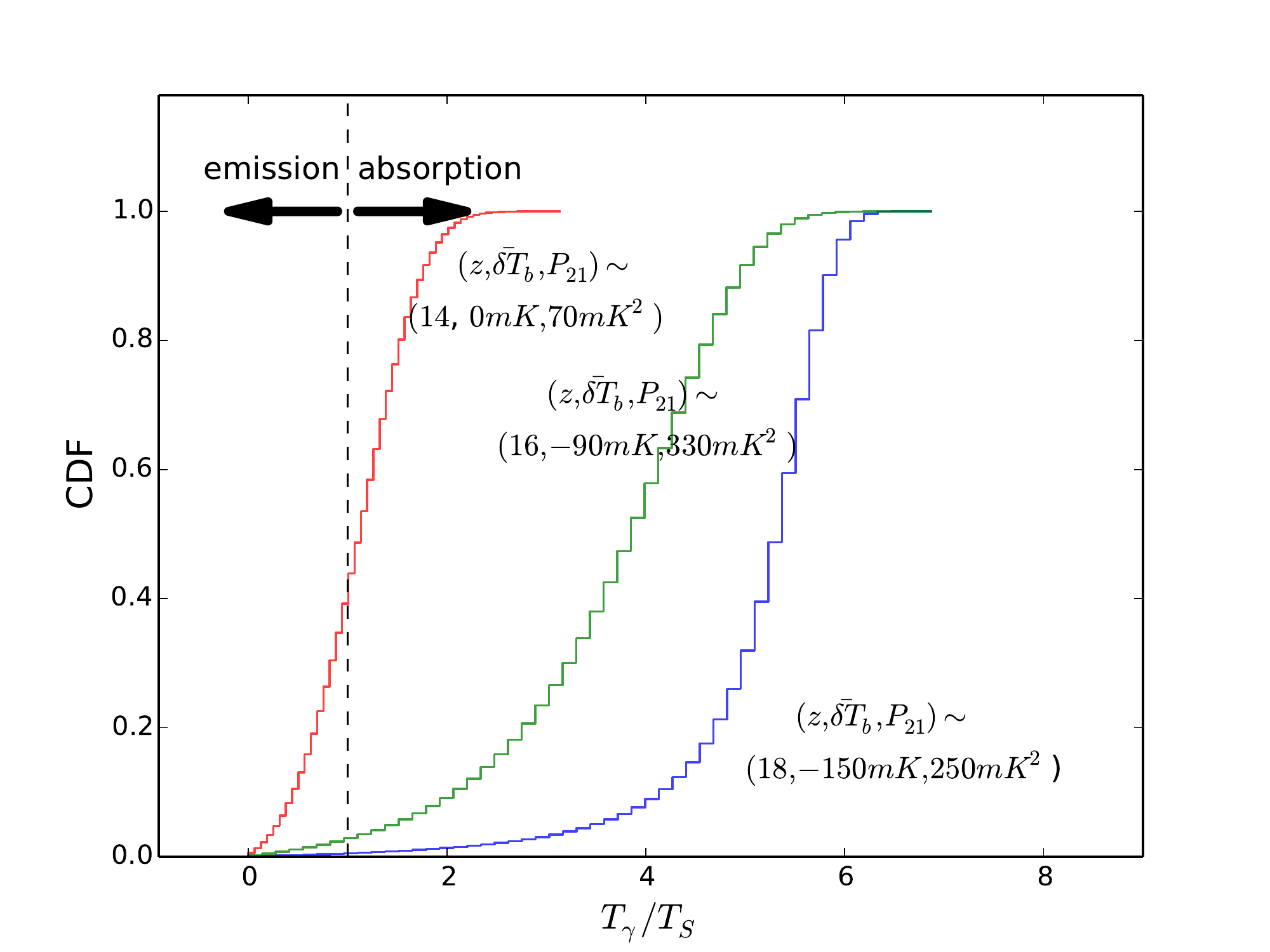}
}
\caption{
Cumulative distributions of $\Tcmb/T_S$ (see eq. \ref{eq:delT}), at $z=14, 16, 18$ (spanning the X-ray peak) for a fiducial, $\Tvir=10^4$ K, $f_X=1$ model.  As seen above, the power peaks when the distribution of $\Tcmb/T_S$ is the broadest.
\label{fig:temp}
}
\vspace{-1\baselineskip}
\end{figure}

The peak power in our parameter space spans the range $30 \lsim P_{21}/$mK$^2 \sim 5000$. However, for a large-swath of reasonable models ($10^{-2} \lsim f_X \lsim 10^2$), the peak power is remarkably constant at a few hundred mK$^2$.
For most models, the power peaks when the mean brightness temperature is $\sim$ -100 mK.  The fluctuations in $\Tcmb/T_S$ are maximized at this time, when areas surrounding sources have been heated to above the CMB temperature, and yet the bulk of the IGM is still cold, seen in absorption.  This can be seen in Fig. \ref{fig:temp}, where we plot the cumulative distribution functions (CDFs) of $\Tcmb/T_S$ at three redshifts spanning the X-ray peak of the fiducial model.  When the amplitude of the power is largest, only a few percent of the IGM is seen in emission (middle curve in Fig. \ref{fig:temp}).  Shortly afterwards, the distribution of $\Tcmb/T_S$ piles up around zero, and the temperature fluctuations cease being important.  Hence, this process is self-similar for many models, especially given that the bias of the halos, which would impact the temperature fluctuations, does not evolve dramatically over our chosen mass range when compared at the same astrophysical epoch\footnote{For fixed astrophysical parameters, such as $f_X$ and $f_{\rm UV}$, a higher value of $\Mmin$ delays the milestones in the signal.  Hence the same astrophysical epoch, such as reionization or X-ray heating, corresponds to a lower redshift when the halo mass function has already evolved.  Instead, when comparing models at the same redshift (and analogously the same mass function), the imprint of the different halo bias resulting from different choices of $\Mmin$ values is more notable.}
(e.g. \citealt{McQuinn07}). Therefore, the lack of notable change in the X-ray peak height is understandable.  However, we caution that the precise peak height and power spectrum evolution is likely affected by the spectral energy distribution of X-ray sources; we postpone an investigation of this to future work.

Understandably, the strongest trend in peak power is with $f_X$; however both extrema are found in models with high values of $\Mmin$.  This is because the growth of structures is both delayed and more rapid in models with higher $\Mmin$; we elaborate more on these trends below.

The weakest signal is found in models with a high $f_X$.  In these models, X-ray heating starts at a high redshift, closely following the onset of Ly$\alpha$ pumping. As a result, the spin temperature does not have time to couple strongly to the kinetic temperature, before the later is driven up by X-ray heating (see the damping of the global absorption trough of the green curve in the bottom panel of Fig. \ref{fig:evolution}). This is clearly evident in lower left panel of Fig. \ref{fig:max_values}: models with high X-ray emissivities have much higher (less negative) values of the mean brightness temperature, $\bar{\delta T}_b$, with the peak power in very high $f_X\sim10^3$ models occurring during the ``emission'' regime (when $\bar{\delta T}_b$ is positive).
 This overlap of WF coupling and X-ray heating increases with increasing $\Mmin$ (or analogously decreasing $\Mwdm$), due to the rapid growth of structures on the high-mass tail of the mass function.  Although not the focus of this study, we note that weaker \lya\ pumping could also result in an overlap of these two epochs.  In particular, if the direct stellar emission was 100 times weaker than assumed, then the 21cm peak power would be decreased by a factor of few for even the fiducial value of $f_X\sim1$.

The strongest signal on the other hand, is found in models with a low $f_X$.  In these models, heating (or reionization) starts later when the $\delT$ contrast is larger due to the evolution of the ratio $\Tcmb/T_S$ (eq. \ref{eq:delT}; note that due to the expansion of the Universe, $\Tcmb/T_S$ evolves roughly as $\propto (1+z)^{-1}$ prior to heating).  Again, the signal is even stronger at high values of $\Mmin$ (or analogously low values of $\Mwdm$), which further shift the evolution to lower redshifts.

The allowed peak amplitude decreases somewhat, when considering only models which are within the 2 $\sigma$ constraints on $\tau_e$ from WMAP9 (demarcated with the solid black curves)\footnote{We compute $\tau_e$ from the evolutions of the average neutral ionized fraction, $\langle x_{i} \rangle$, and density, $\langle n \rangle$.  Strictly speaking, the correlations between these two fields should be taken into account, i.e.  $\langle x_{i} \times n \rangle$ $\neq$ $\langle x_{i} \rangle \times \langle n \rangle$.  \citet{MFS13} note that the fact that reionization is ``inside-out'' on large scales results in a slightly higher value of $\tau_e$, than estimated ignoring correlations.  We account for this bias by multiplying our $\tau_e$ estimates by 1.04, the size of the bias in fiducial, UV driven reionization scenarios. We further assume that reionization has completed by our last redshift output at $z=7$ (i.e. imposing $\avenf=0$ at $z<7$).  This means that $\tau_e$ is overestimated somewhat for models with high $\Mmin \gsim 10^9 \Msun$ which do not complete reionization by this redshift.  Hence the upper 2 $\sigma$ contour is conservatively broad.}.  Limits on $\tau_e$ generally rule out late reionization models, which for our choice of $\zeta_{\rm UV}$ correspond to high values of $\Mmin \gsim 10^{9.5}\Msun$.  However, at efficiencies $f_X\gsim10$, X-rays begin to contribute to reionization at the $\gsim10$\% level (c.f. \citealt{MO12, MFS13}), and the WMAP9 $\tau_e$ isocontours curve upward, including high $\Mmin$ models.

We can also see from the lower right panel in Fig. \ref{fig:max_values} that for $f_X \gsim 10^{-2}$ the Universe is mostly neutral when the 21cm power peaks. This means that the peak power indeed occurs during the X-ray heating epoch, before reionization.  In contrast, the reionization peak in power (c.f.. top panel of Fig. \ref{fig:evolution}), should occur at $\avenf\sim0.5$ (e.g. \citealt{Lidz08, Friedrich11, MFS13}).  In our models, this occurs for values of $f_X \lsim 10^{-2}$, consistent with simple analytic estimates \citep{MO12}.  In these models, the X-ray background is too weak to heat the IGM before reionization.  The resulting contrast between the (very) cold neutral and ionized patches drives the reionization power to values of $\gsim10^3$ mK$^{2}$ (c.f. \citealt{Parsons13}).

\subsection{Detectability of the peak power}

\begin{figure*}
\vspace{-1\baselineskip}
{
\includegraphics[width=0.33\textwidth]{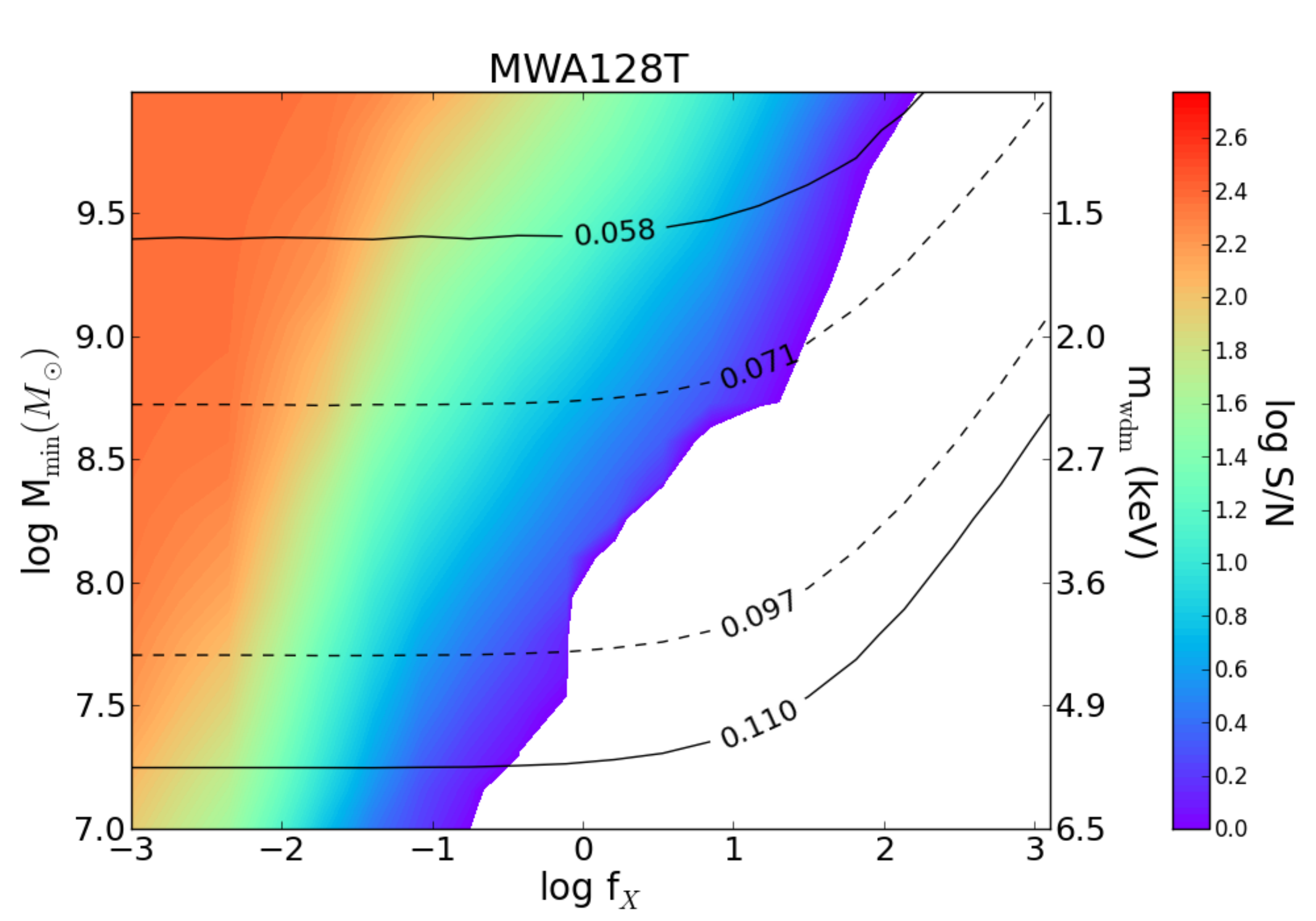}
\includegraphics[width=0.33\textwidth]{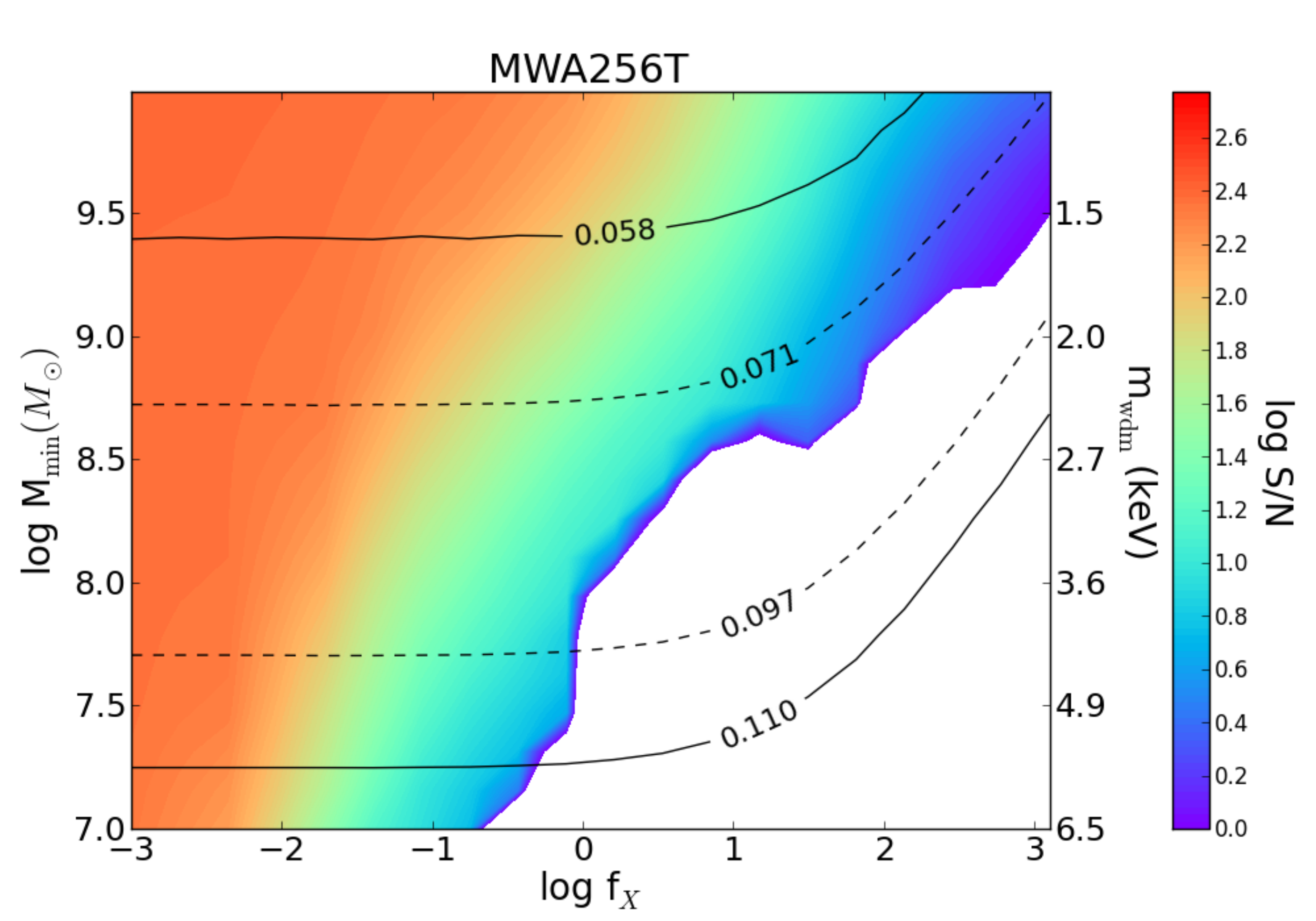}
\includegraphics[width=0.33\textwidth]{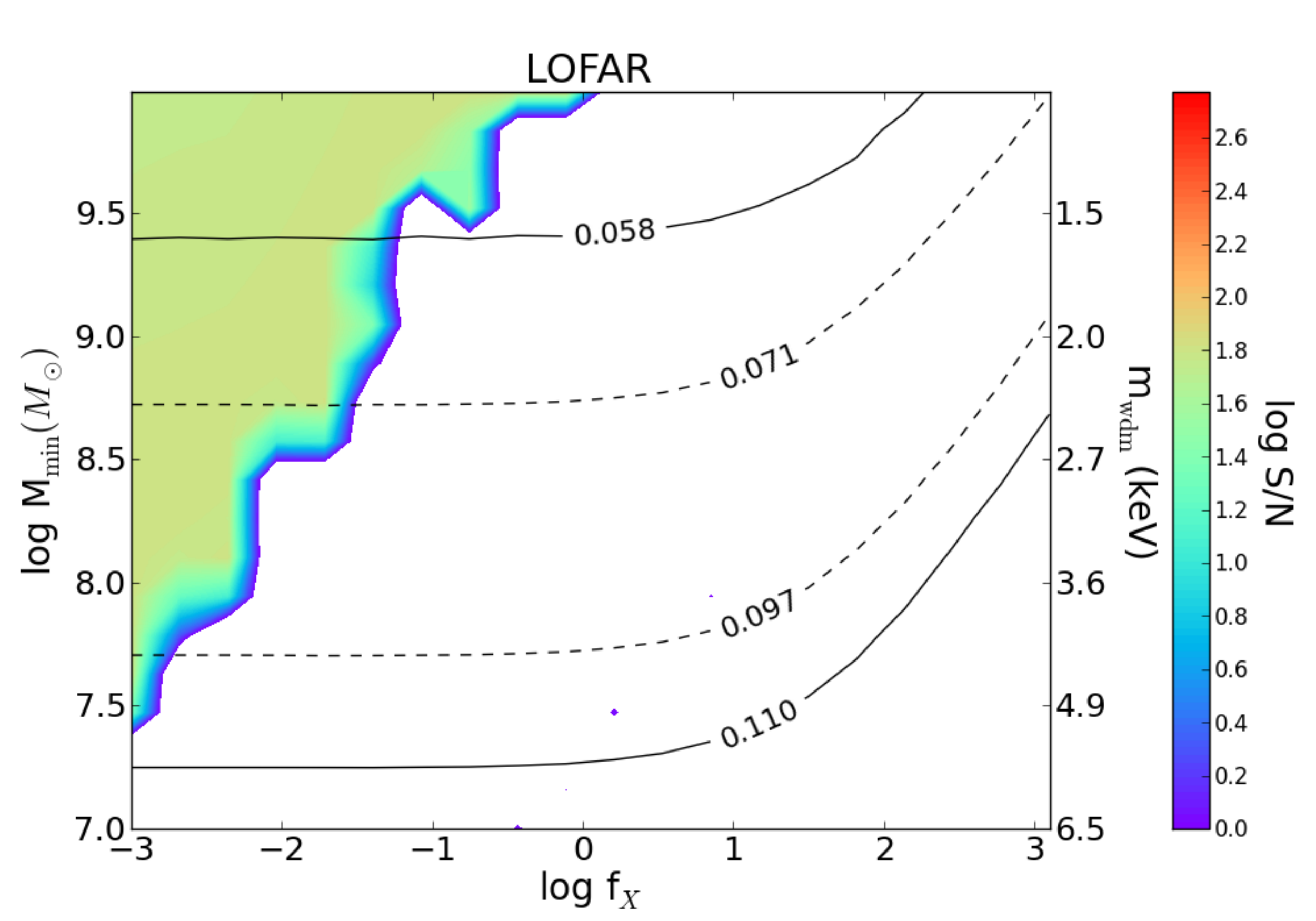}
\includegraphics[width=0.33\textwidth]{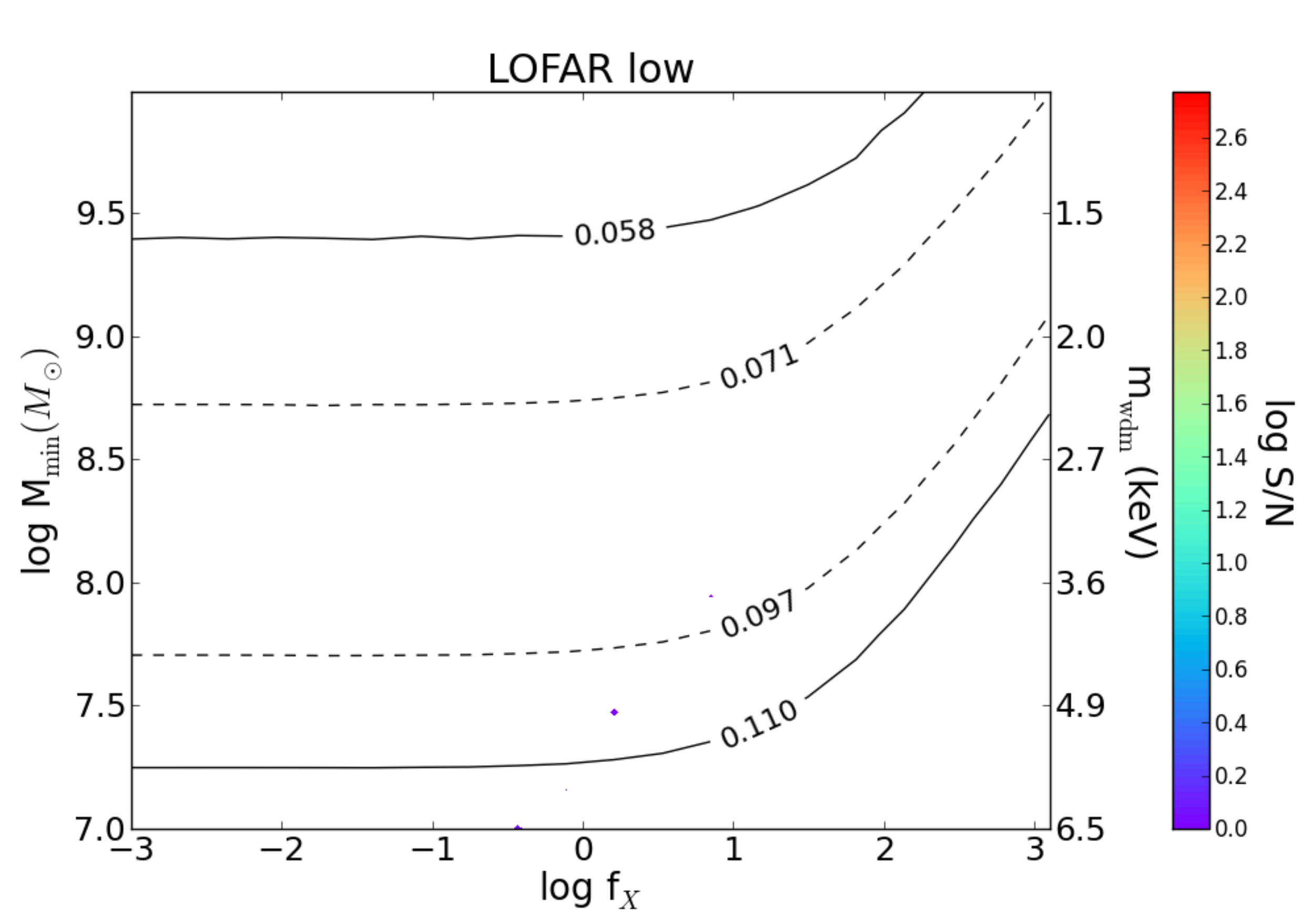}
\includegraphics[width=0.33\textwidth]{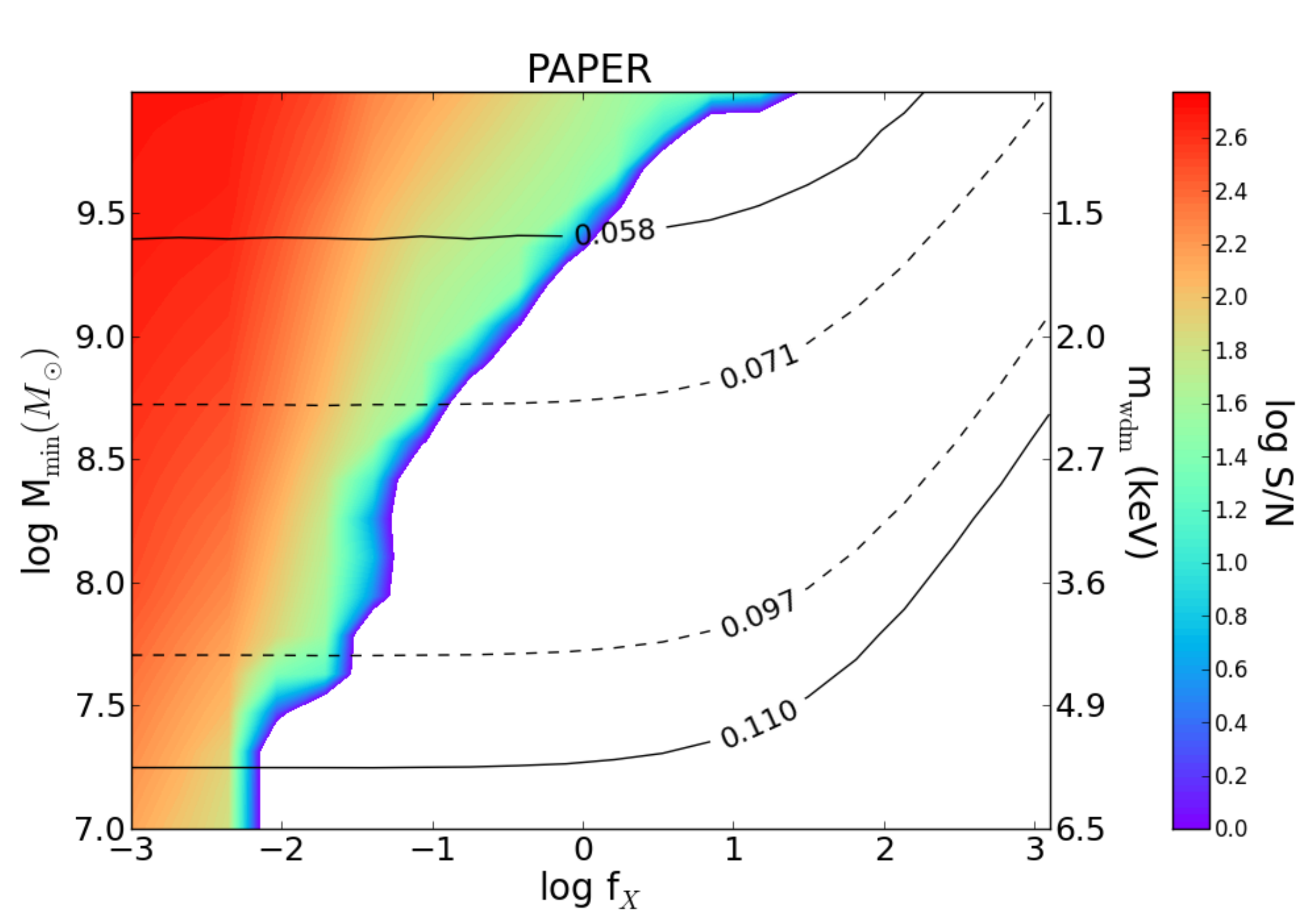}
\includegraphics[width=0.33\textwidth]{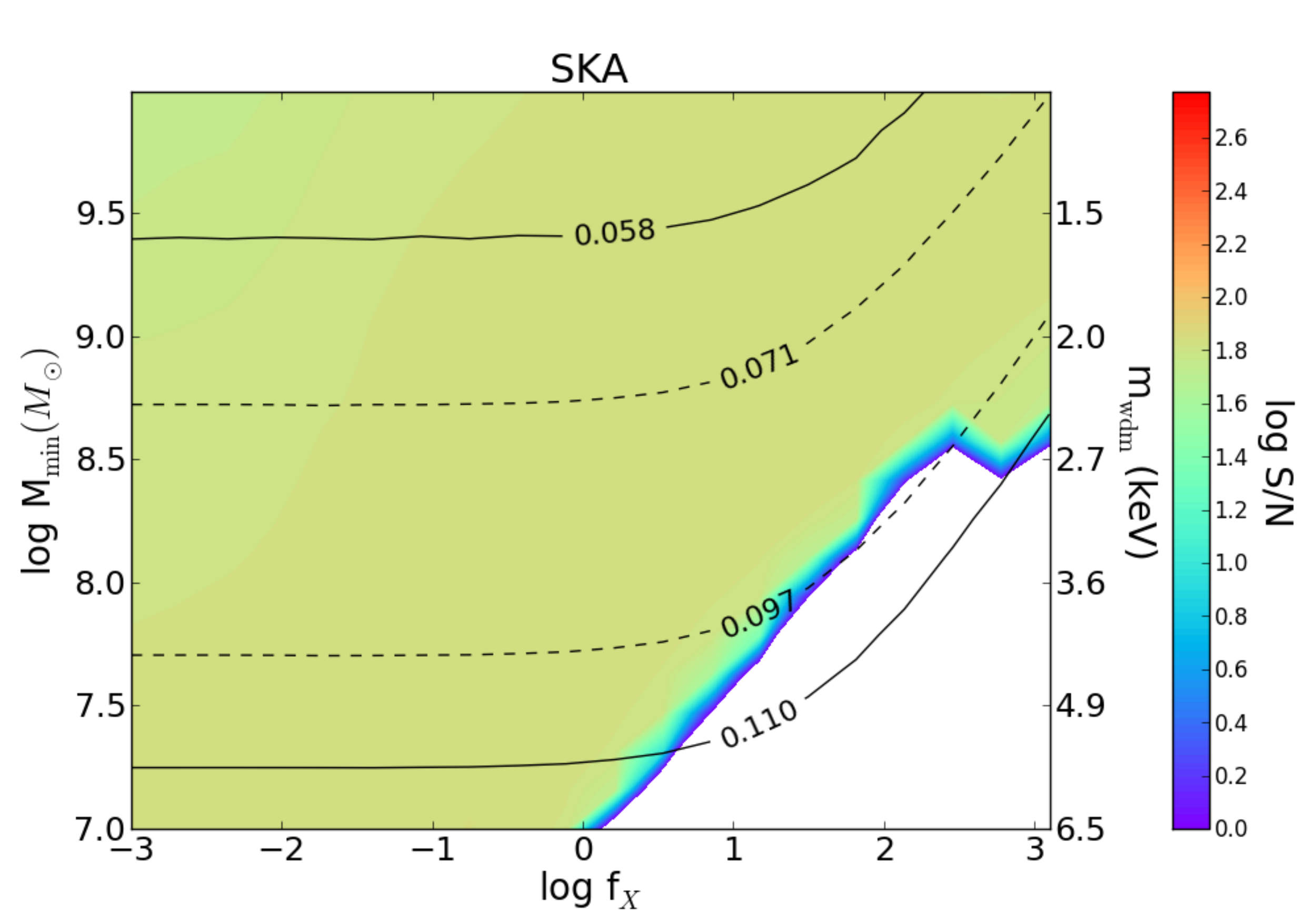}
}
\caption{
S/N of the detection of the $k=0.1$ Mpc$^{-1}$ 21cm peak power (i.e. computed at the redshift of the maximum signal).  All maps show the same range in S/N to highlight differences between instruments, and are computed assuming a total integration time of 2000h.  Due to our fiducial observing strategy which minimizes thermal noise at the expense of cosmic variance, some high S/N regions (S/N $\gsim 50$) are limited by cosmic (Poisson) variance for the case of LOFAR and SKA, which have smaller beams than MWA and PAPER (see Fig. \ref{fig:poisson}).  This Poisson variance limit can be avoided with a different observing strategy; hence we caution the reader not to lend weight to the apparent better performance of MWA and PAPER in the upper left region of parameter space.
\label{fig:sigma}
}
\vspace{-0.5\baselineskip}
\end{figure*}

We now include our 2000h sensitivity estimates from \S \ref{sec:sens}, in order to predict the detectability of the X-ray heating peak with current and future interferometers.  In Fig. \ref{fig:sigma}, we plot the S/N according to eq. (\ref{eq:SN}), with which the peak power can be detected.  Regions in white correspond to areas of parameter space where the signal to noise is less than unity.

The isocontours of S/N generally follow the diagonal trend of the redshift isocontours from the top right panel in Fig. \ref{fig:max_values}.  This is due to the fact the sensitivities of the interferometers (noise) vary more strongly with redshift than does the amplitude of the peak power (signal).  The exception to this trend is the strip at $f_X \lsim 10^{-2}$, corresponding to reionization in a cold IGM, when the power can jump to $P_{21} \gsim 10^3$ mK$^2$.  This ``cold-reionization peak'' is strongly detectable by all instruments.  Therefore, it should be noted that {\it there is no such thing as an ``X-ray heating peak'' for $f_X \lsim 10^{-2}$}.

The peak power of a fiducial model ($\Mmin\approx10^8 \Msun$, $f_X\approx1$) lies at the edge of detectability for MWA128T.  However, if X-ray heating is delayed, either by lowering $f_X$ or increasing $\Mmin$ by a factor of $\sim10$, 128T MWA can detect the X-ray peak at the S/N $\sim$ few--10 level.  

The bandpass coverage means that the parameter space region in which S/N is greater than unity (at peak signal) does not evolve much when upgrading the MWA to 256 tiles.  However, the S/N of a detection increases dramatically for MWA256, with the X-ray peak being detectable at the S/N $\gsim$ 10 level throughout the bandpass.


In the case of LOFAR, we find that the peak signal is detectable at S/N $> 1$ only with the high-band antennas (we keep the blank LOFAR low panel in Fig. \ref{fig:sigma} for the sake of symmetry).  Only late heating models are therefore detectable, corresponding to the high-$\Mmin$, low-$f_X$ corner\footnote{The lower S/N of these detections compared with the analogous ones by the MWA results from the Poisson (cosmic variance) limit (see Fig. \ref{fig:poisson}).  LOFAR and SKA have a narrower beam than the MWA.}.  If one discounts the ``cold-reionization'' $f_X\lsim10^{-2}$ strip, as well as the low-$\tau_e$ region, it is apparent that LOFAR is unlikely to detect the peak power during X-ray heating. 
However, even when the actual peak signal lies beyond the band pass, it is possible that the X-ray heating power might still extend into the LOFAR low bandpass, allowing for low S/N detections (see for example the fiducial model in Fig. \ref{fig:evolution} in which the power peaks outside the bandpass at $z\sim16$, but is still detectable at $z\sim17$). We will quantify this below when we plot the peak S/N for LOFAR low.

The results for PAPER are quite similar to those of LOFAR.  However, the slightly wider PAPER bandpass extending to lower frequencies allows for a slightly wider strip of X-ray heating detections in Fig. \ref{fig:sigma}. 

Finally, it is evident from the bottom right panel of Fig. \ref{fig:sigma} that the second generation interferometer, SKA, will be a huge improvement over the first generation instruments.  SKA should be able to detect all X-ray heating models we consider, with the exception of the lower right corner in which the peak signal extends beyond our fiducial coverage, $z \gsim 20$ (we remind the reader that this is not a fundamental limitation of SKA, and can be avoided with a wider bandwidth choice).  In fact, all of the SKA detections are Poisson noise limited (c.f. Fig. \ref{fig:poisson}).  Our observational strategy was chosen to minimize thermal noise at the expense of Poisson noise.  If however even stronger detections are desired with the SKA, one could observe more fields for a shorter period of time (however the benefits of a S/N $\gsim 500$ detection compared with a S/N $\gsim50$ one, are not immediately obvious).

\subsection{General detectability of the cosmic 21cm signal}
\label{sec:gen_dect}

\begin{figure*}
\vspace{-1\baselineskip}
{
\includegraphics[width=0.33\textwidth]{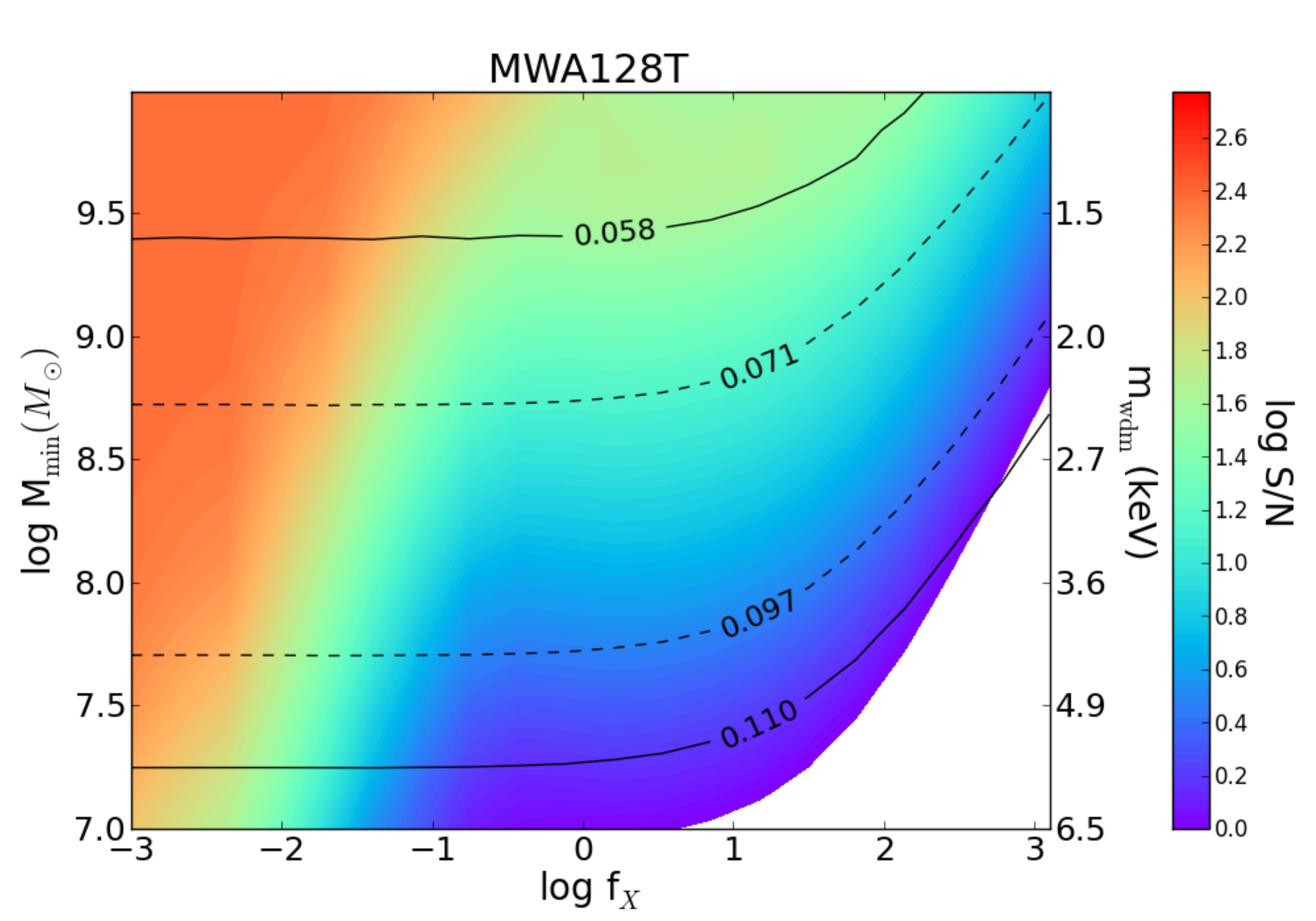}
\includegraphics[width=0.33\textwidth]{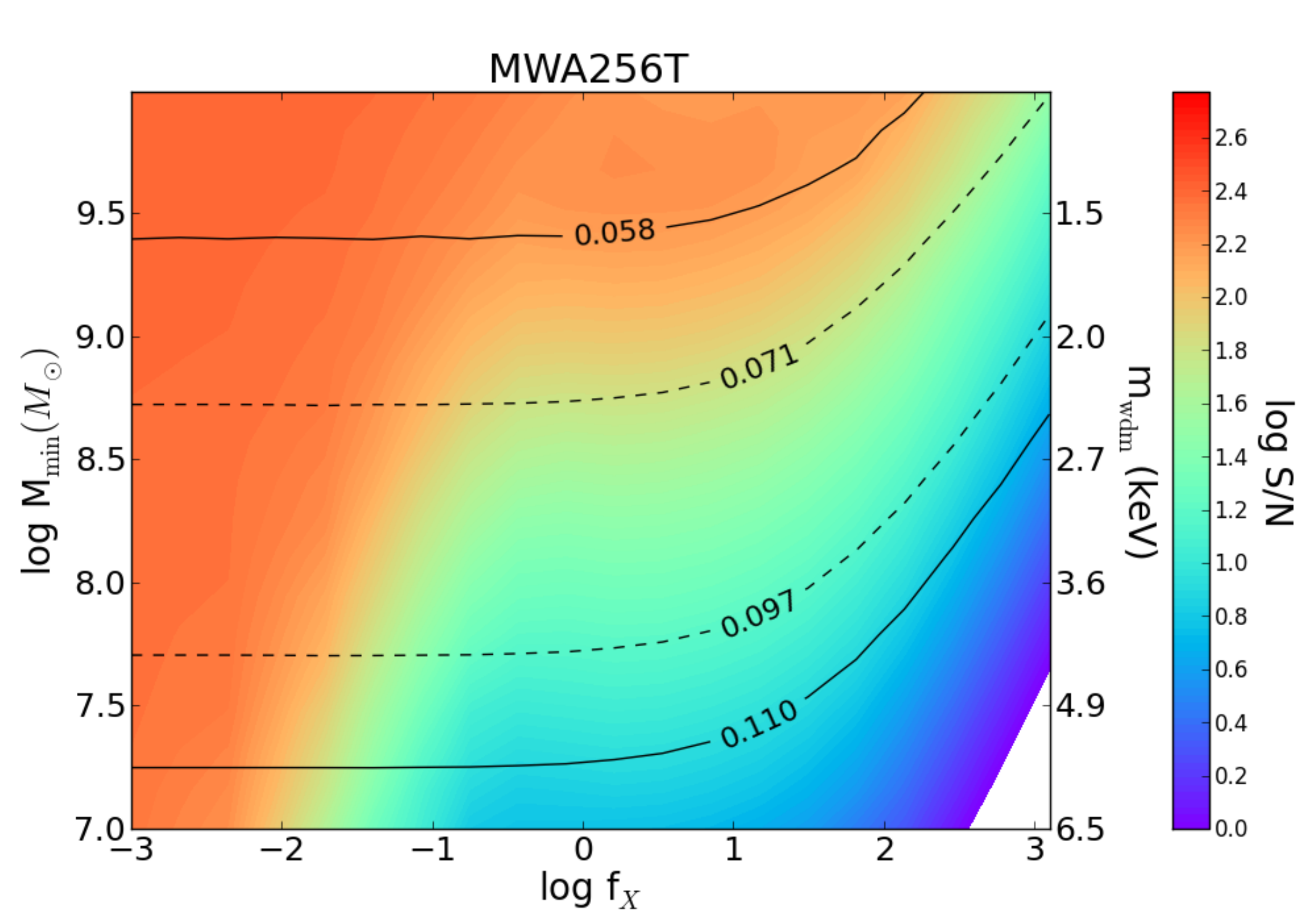}
\includegraphics[width=0.33\textwidth]{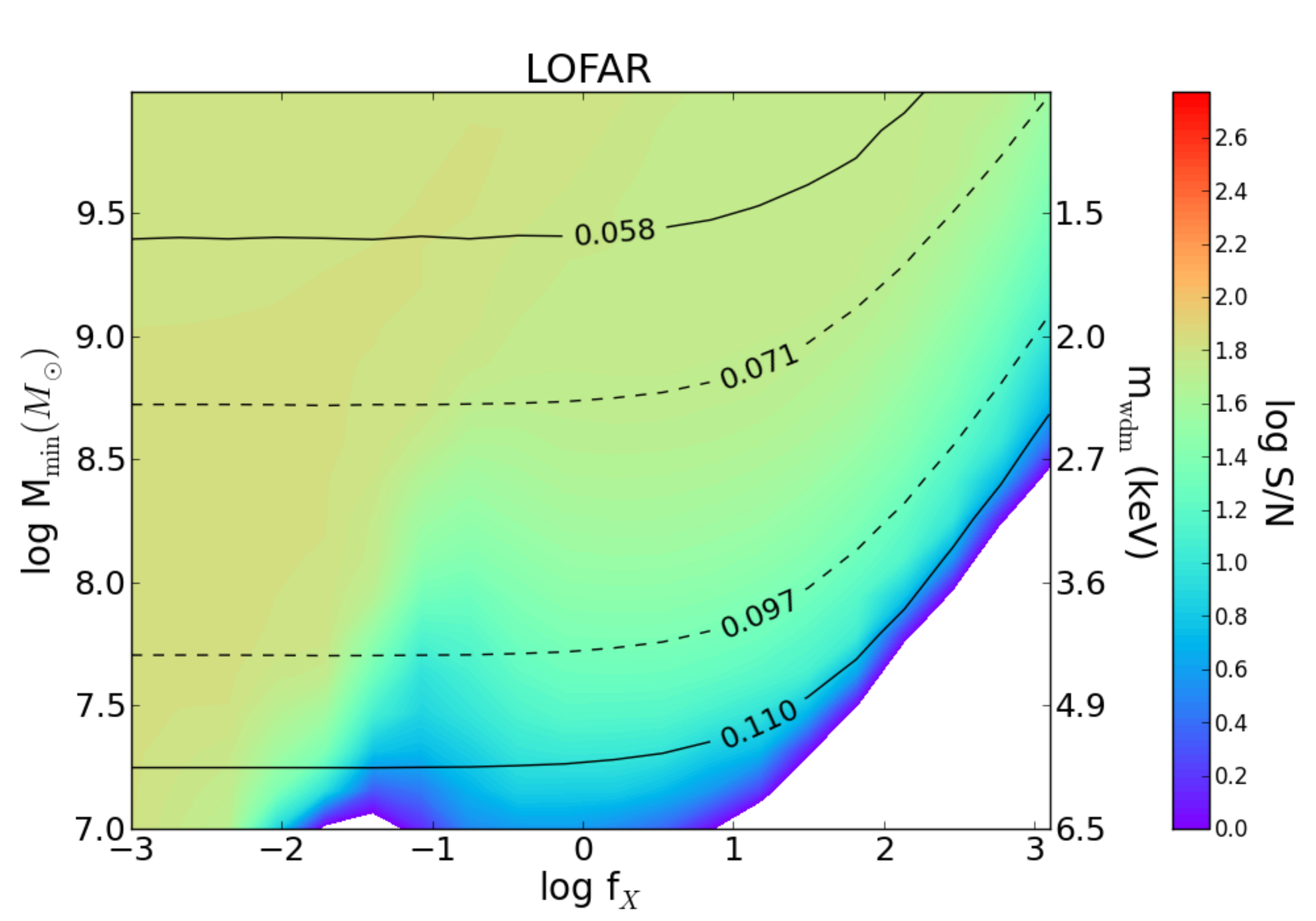}
\includegraphics[width=0.33\textwidth]{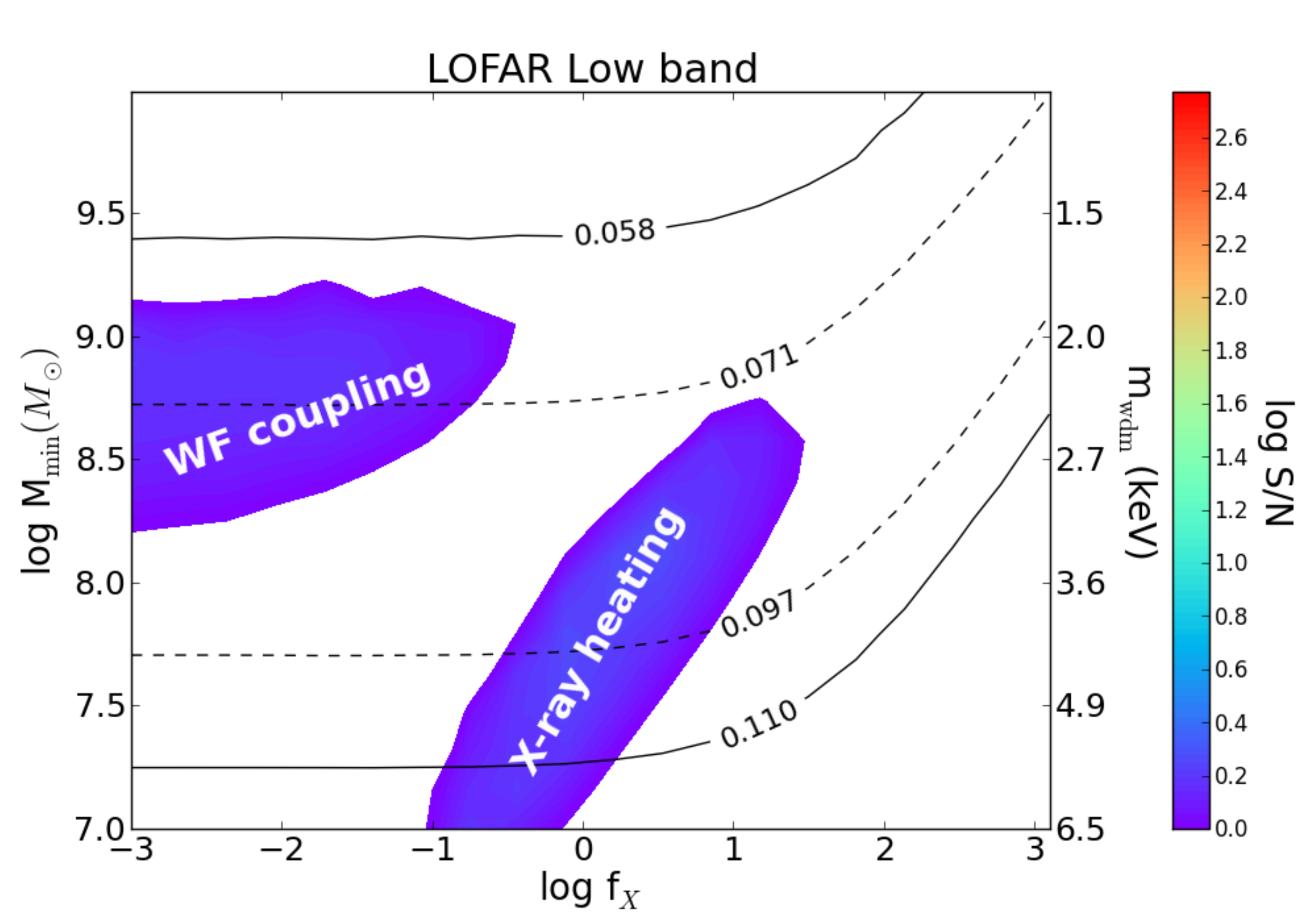}
\includegraphics[width=0.33\textwidth]{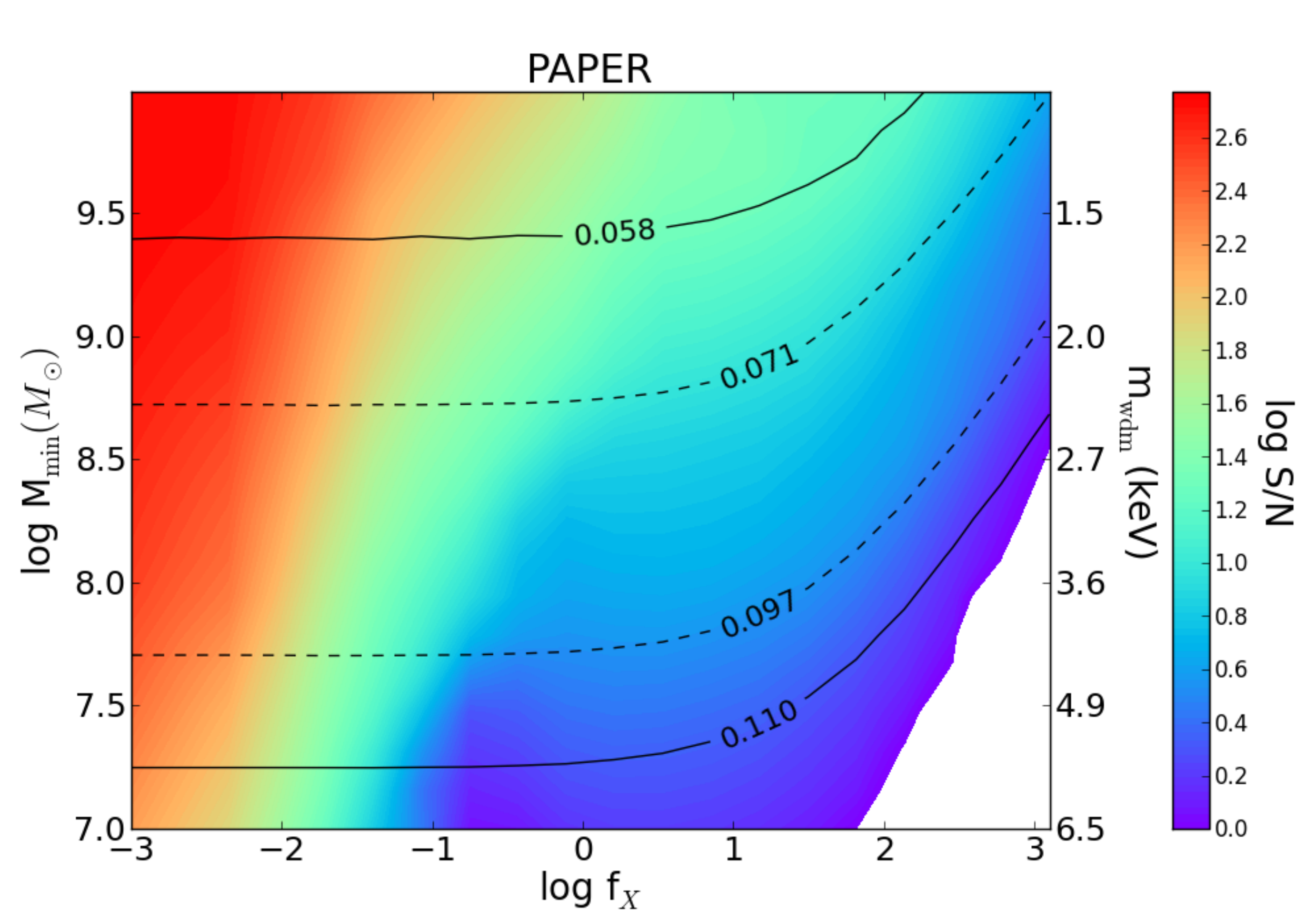}
\includegraphics[width=0.33\textwidth]{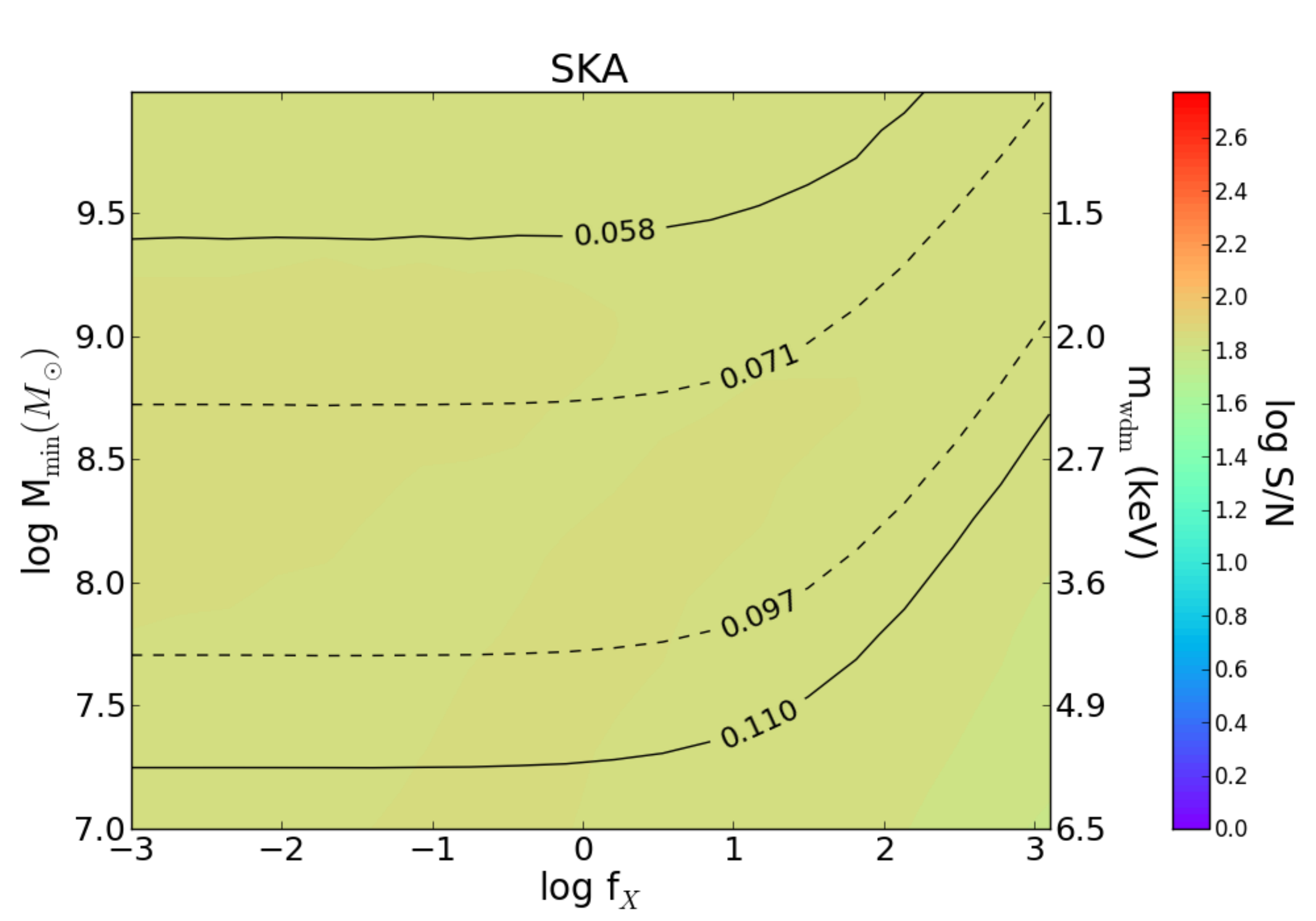}
}
\caption{
Maximum S/N possible with current and upcoming interferometers after 2000h (considering all redshifts instead of just the peak signal as in Fig. \ref{fig:sigma}).
\label{fig:sigma_maxSN}
}
\vspace{-0.5\baselineskip}
\end{figure*}

Thus far we have focused on the maximum amplitude (over all redshifts) of the 21cm power spectrum at $k=0.1$ Mpc$^{-1}$.  This peak power generally corresponds to the epoch of X-ray heating ($f_X \gsim 10^{-2}$), or ``cold-reionization'' ($f_X \lsim 10^{-2}$).  We now ask instead how detectable is the cosmic 21cm signal, {\it regardless of the epoch}.  Due to the increase in instrument noise towards higher redshifts, in many models the reionization peak is more detectable than the X-ray one, even though the cosmic signal is weaker (see Fig. \ref{fig:evolution}).  

In Fig. \ref{fig:sigma_maxSN}, we show S/N plots as in Fig. \ref{fig:sigma}, but computed at the maximum S/N (instead of the maximum signal). Understandably, the detectable region of parameter space broadens for all instruments, as models in which the peak signal fell out of the observable bands are now again considered. This broadening of detectable parameter space is most dramatic for LOFAR and PAPER, whose band passes are optimized for reionization.

As mentioned above, we see that LOFAR low could catch the pre-reionization signal (albeit at S/N $\sim$ unity), even though it could not detect the actual peak of the signal.  Indeed our ``fiducial model'' (c.f. the black curve in the left panel of Fig. \ref{fig:evolution}) falls into this category, with the {\it pre-}peak power extending into the LOFAR-low detection region.  It is also interesting to note that the upper strip in the LOFAR low panel of Fig. \ref{fig:sigma_maxSN} corresponds to the earlier Ly$\alpha$ pumping peak, before X-ray heating, which is also detectable at S/N$\sim$ unity.  This strip includes models in which the X-ray heating is sufficiently delayed to allow the Ly$\alpha$ pumping epoch to extend into the LOFAR-low detection region (c.f. the blue curve in the left panel of Fig. \ref{fig:evolution}).

\begin{figure}
\vspace{-1\baselineskip}
{
\includegraphics[width=0.5\textwidth]{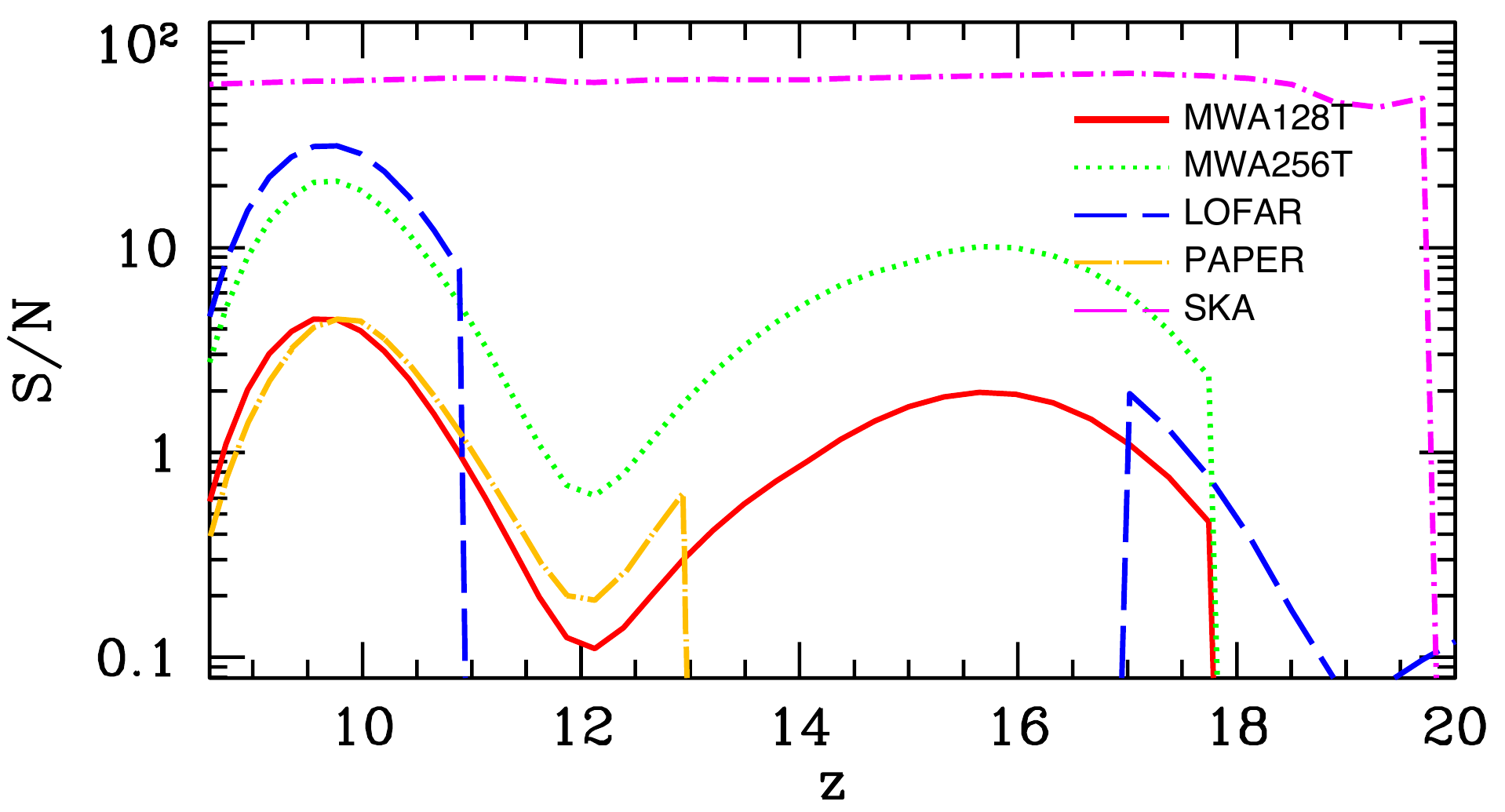}
}
\caption{
The S/N vs redshift evolution for the ``fiducial'' model, with $\Mmin=10^8 \Msun$, $f_X=1$.
\label{fig:fidSN}
}
\vspace{-1\baselineskip}
\end{figure}

In Fig. \ref{fig:fidSN}, we show the evolution of the S/N for the ``fiducial'' model: $\Mmin=10^8 \Msun$, $f_X=1$.  We can see that reionization is detectable at $\approx 4\sigma$ with MWA128T
 and PAPER, and at $\approx$20--30$\sigma$ with LOFAR and the potential MWA256T.  MWA128T and LOFAR might be able to detect X-ray heating in the fiducial model at $\approx$1--2$\sigma$.  The SKA is cosmic variance limited throughout.

It should be noted that in this work, we compute the S/N from a single $k$-bin and frequency (i.e. redshift) bin.  In principle, one can boost the S/N of the detection by a weighted sum over both the available $k$-modes as well the available frequency bins.  Depending on the foreground smoothness, the first generation instruments might only have a narrow window in $k$-space to make the measurement.  However, the signal could still have a S/N $\gsim 1$ over several frequency (redshift) bins (see Fig. \ref{fig:fidSN}).  Hence, by summing over these bins, the S/N of the detection could be boosted by an additional factor of $\sim$ few, provided there is contiguous frequency coverage surrounding the relevant epochs.


It is important to also note that for continuous and relatively wide band pass interferometers like the MWA, the change between the S/N computed at max signal vs that computed at max S/N (Fig. \ref{fig:sigma} vs Fig. \ref{fig:sigma_maxSN}) is not very dramatic: mostly low S/N detections extend into the parameter space where the peak X-ray heating signal occurs too early to make MWA's $z\approx18$ band cut.   We can also see this explicitly for the fiducial model in Fig. \ref{fig:fidSN}: the S/N of X-ray heating is only a factor of $\approx2$ less than that of reionization for MWA128T.
 This is suggestive that for these instruments, the detectability of reionization and X-ray heating are roughly comparable from a thermal noise perspective \footnote{Additional observational challenges associated with the lower frequencies of the X-ray heating epoch such as calibration and RFI are not considered in this work.}.  We explore this in the following section.

\subsection{Reionization or X-ray heating?}
\label{sec:rvsx}

We can now ask the question: ``does the maximum S/N correspond to the reionization or the X-ray heating epoch?''  Aside from the bandpass limitations of LOFAR and PAPER, this distinction is not so clear.

\begin{figure*}
\vspace{-1\baselineskip}
{
\includegraphics[width=0.33\textwidth]{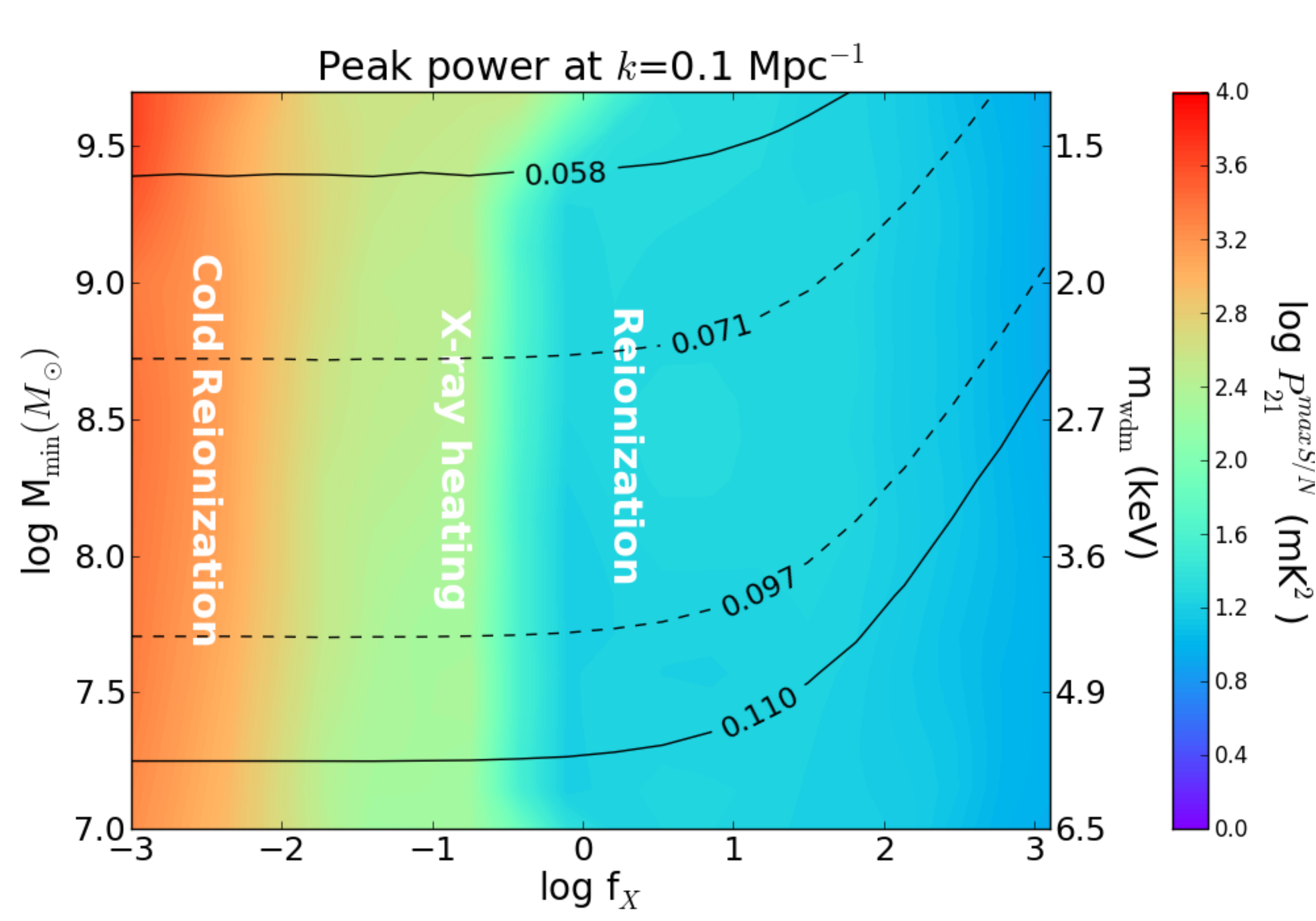}
\includegraphics[width=0.33\textwidth]{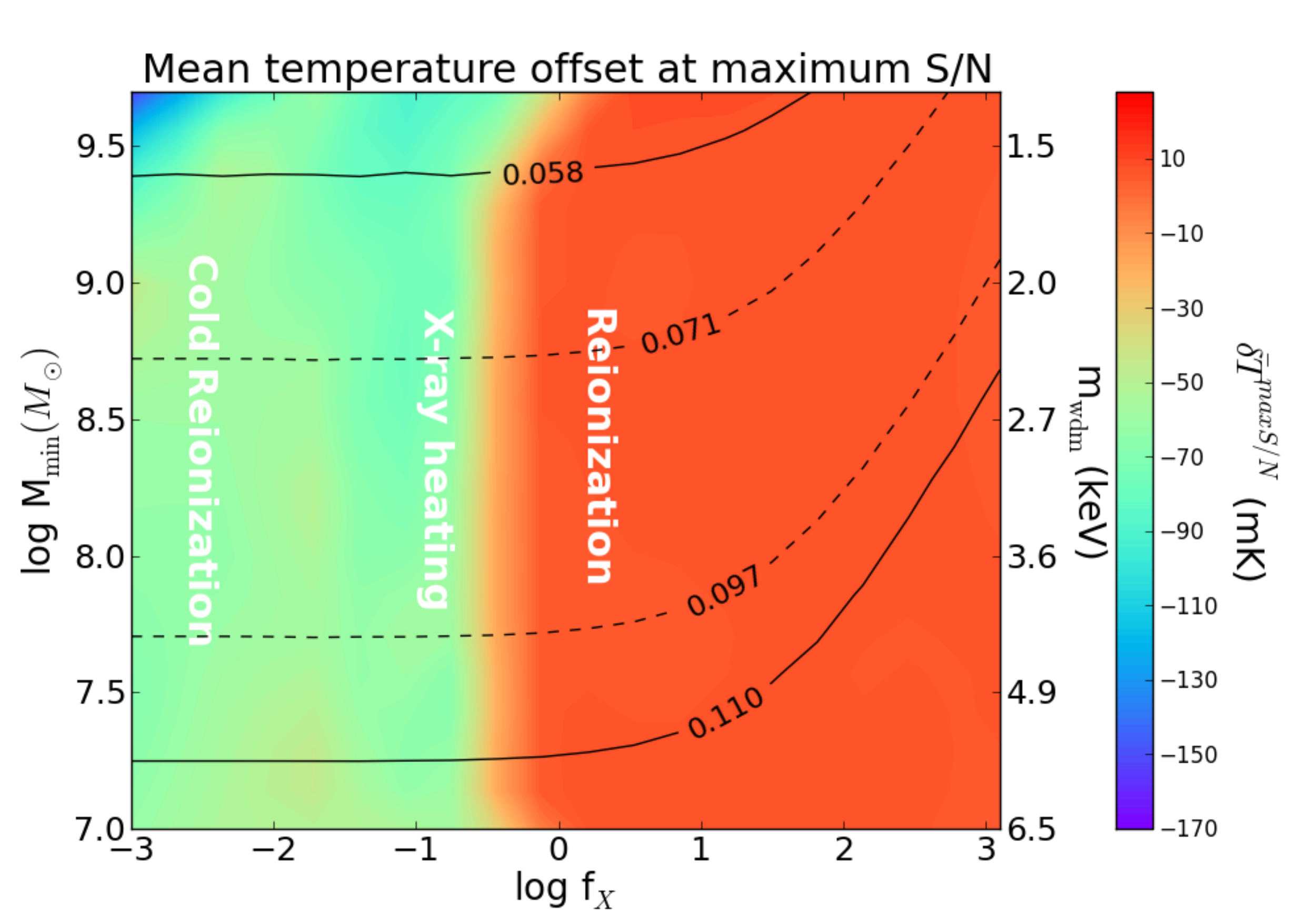}
\includegraphics[width=0.33\textwidth]{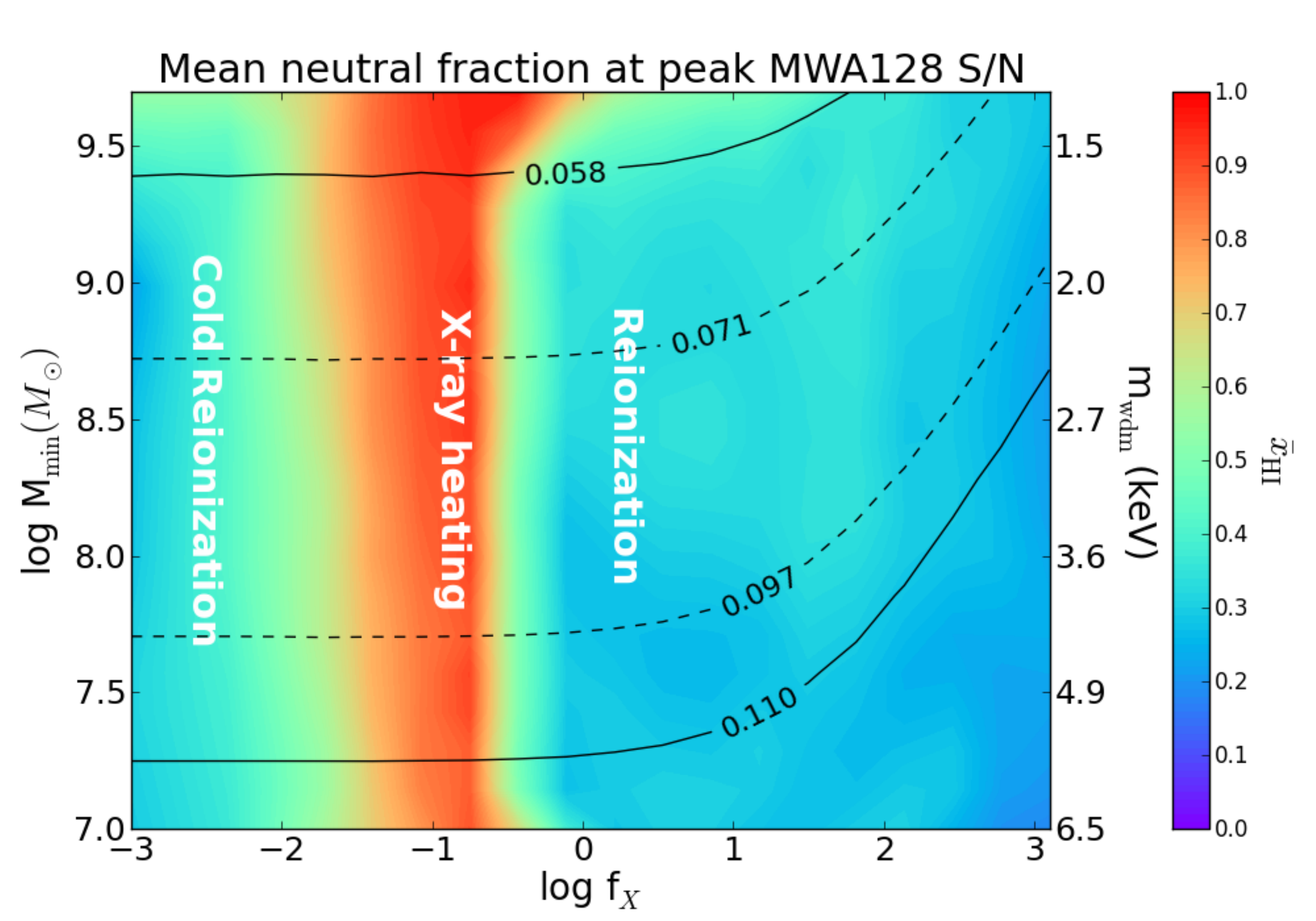}
}
\caption{
The 21cm power spectrum amplitude at $k=0.1$ Mpc$^{-1}$ ({\it left}), $\bar{\delT}$ ({\it center}), and $\avenf$ ({\it right}), all evaluated at the redshift when S/N is the largest, assuming MWA128 sensitivities.
\label{fig:MWApsxH}
}
\vspace{-1\baselineskip}
\end{figure*}

To further quantify this, in Fig. \ref{fig:MWApsxH} we plot the 21cm power spectrum amplitude ({\it left}),  $\bar{\delT}$ ({\it center}), and the neutral fraction ({\it right}), all computed at the redshift of peak S/N, assuming MWA128T sensitivities. Although we do not include the plots, we note that the trends for continuous frequency coverage instruments like the SKA are the same, in the regime when the detections are not cosmic variance limited (i.e. excluding the Poisson contribution to the noise).
  From Fig. \ref{fig:MWApsxH}, we see that there is a clear separation between the regimes where the peak S/N is achived during reionization vs during X-ray heating.

As already noted, during the reionization epoch (more precisely, when the power spectrum of the ionization field is dominating the total 21cm signal), the large-scale 21cm power peaks at the mid-point of reionization (e.g. \citealt{Lidz08, Friedrich11}), except for extreme models with $f_X>10^3$ \citep{MFS13}.  The peak of $k=0.1$ Mpc$^{-1}$ power during reionization is roughly $\sim10$ mK$^2$, more than an order of magnitude less than the peak during the X-ray heating epoch.  From Fig. \ref{fig:MWApsxH}, we see that  models with $f_X \gsim 1$ indeed follow these reionization epoch predictions\footnote{If X-rays contribute significantly to reionization (e.g. very high values of $f_X$), then the large-scale power during the advanced stages of reionization can peak at lower neutral fractions (later stages).  This is due to the suppression of ionization power from X-rays, whose mean free paths result in a smoother reionization \citep{MFS13}. As reionization progresses however, a smaller fraction of the photon energy goes into ionizations, and the relative contribution of UV photons increases.}.

  Instead for $10^{-1.5} \lsim f_X\lsim 10^{-0.5}$, the Universe is mostly neutral at the highest S/N, with peak power amplitudes of $\gsim100$ mK$^2$.  This corresponds to the epoch of X-ray heating, signifying that for $10^{-1.5} \lsim f_X\lsim 10^{-0.5}$ the X-ray peak becomes more detectable than the reionization peak.
This is understandable by considering the positions of the reionization and X-ray heating peaks.  We can already see from Fig. \ref{fig:evolution}, that for a fiducial model ($f_X\sim1$), the increase in the interferometer noise between reionization and X-ray heating epochs is almost (not quite) compensated by the increase in the signal.  As the X-ray efficiency is lowered, the heating epoch moves to lower redshifts and becomes more detectable than the reionization epoch.  

As the X-ray efficiency is lowered even further, $f_X \lsim 10^{-2}$, we enter into the regime of ``cold reionization'': with X-rays unable to heat the IGM before the completion of reionization.  The corresponding contrast between the cold neutral and ionized IGM drives the 21cm power to amplitudes in excess of $10^3$ mK$^2$.  In this regime, there is effectively no ``X-ray heating epoch'', and the highest S/N is again achieved during reionization.

\section{Considering more pessimistic foregrounds}
\label{sec:unsmooth}

 Preliminary observations by \citet{Pober13} suggest that our fiducial choice of $k = 0.1$ Mpc$^{-1}$ is reasonably free of foregrounds, though slightly larger scales lie securely in the ``wedge''.  Here we briefly consider a more pessimistic scenario, in which the frequency structure in the foregrounds contaminates modes even out to $k = 0.2$ Mpc$^{-1}$.

 Generally speaking, the resulting noise levels at $k = 0.2$ Mpc$^{-1}$ for our sensitivity curves are approximately five times higher than at $k=0.1$ Mpc$^{-1}$ for all instruments considered.  Since the shape of the 21cm power spectrum is flat at the peak amplitude ($dP_{21}/d\ln k \approx 0$; \citealt{MFS13}), we expect the S/N in the thermal noise dominated regime to be a factor of five times less at $k = 0.2$ Mpc$^{-1}$ than at $k = 0.1$ Mpc$^{-1}$.

\begin{figure}
\includegraphics[width=.5\textwidth]{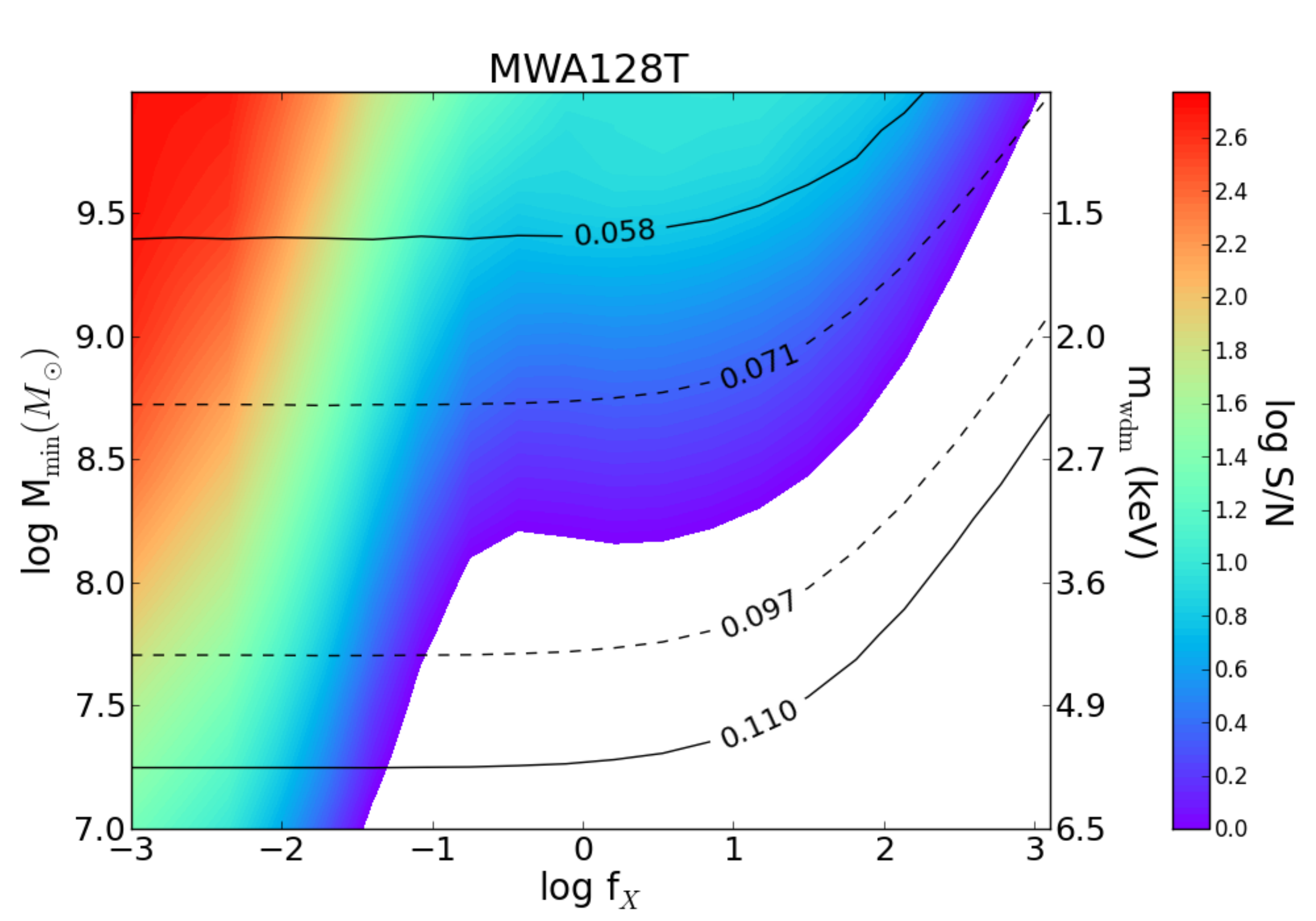}
\caption{The maximum S/N for MWA-128T at $k=0.2$Mpc$^{-1}$.
\label{fig:unsmooth}}
\end{figure}

 In Figure \ref{fig:unsmooth} we show the maximum S/N obtainable with MWA-128T at $k=0.2$ Mpc$^{-1}$ (to be compared with the top left panel of Fig. \ref{fig:sigma_maxSN}). We see that the detectable parameter space shrinks, and the broad S/N $\gsim$ 10 region is now only marginally detectable at S/N $\sim$ unity.
This confirms that we have a relatively narrow $k$-space ``window'' with first generation instruments;  their detection of X-ray heating and reionization might depend on our ability to avoid or mitigate foregrounds at $k \approx 0.1$ Mpc$^{-1}$.

\section{Conclusions}
\label{sec:conc}

Upcoming statistical and (eventual) tomographical studies of the early Universe with 21cm interferometry should dramatically increase our understanding of both astrophysics and cosmology.  The potential of this probe has not yet been fully explored.

Here we perform an astrophysical parameter study, exploring different minimum DM halo masses required to host galaxies, $\Mmin$, as well as the galactic X-ray emissivity, normalized to present-day values, $f_X$. We also discuss the signal in terms of popular warm dark matter models, recasting $\Mmin$ to an analogous warm dark matter particle mass, $\Mwdm$. We study the detectability of the 21cm power spectrum with current and upcoming interferometers.  We quantify the peak S/N as well as the S/N at the peak signal, generally corresponding to the epoch of X-ray heating.

Our two free parameters, $\Mmin$ and $f_X$, are fundamental in controlling the timing and relative offset of reionization and X-ray heating.  The resulting plane of models in parameter space spans the maximum variation in the S/N.  Hence, although we focus on the under-appreciated X-ray heating epoch, we also expect our predictions for reionization and the overall achievable S/N in Fig. \ref{fig:sigma_maxSN} to be robust.

For values of $10^{-2} \lsim f_X \lsim 10^2$, the peak amplitude of the 21cm power at $k=0.1$ Mpc$^{-1}$ is roughly constant at a few hundred mK$^2$.  In this regime, the peak power occurs during X-ray heating, when just a few percent of the IGM is in emission and the fluctuations in $\Tcmb/T_S$ are maximized.
 Stronger X-ray backgrounds instead heat the IGM before its spin temperature has efficiently coupled to the gas temperature, resulting in a weak signal, $P_{21}\sim10$ mK$^2$.  On the other hand, weaker X-ray backgrounds are insufficient to heat the IGM before the completion of reionization.  Reionization in these scenarios proceeds in a cold IGM, with the resulting contrast driving the 21cm power to values in excess of thousands of mK$^2$.

Aside from bandpass limitations,  in ``reasonable'' models (within an order of magnitude of fiducial values) X-ray heating is detectable at roughly comparable signal-to-noise to reionization.  The increase in the signal during the X-ray heating epoch can approximately compensate for the increase in thermal noise going to lower frequencies.  A stronger detection is achievable if X-ray heating occurred late, driven by either X-ray faint galaxies ($f_X\lsim1$) or those hosted by halos more massive than the cooling threshold, $\Mmin > 10^{8} \Msun$ (or analogously if $\Mwdm < 3.6$ keV).

For reasonable models, it is unclear if all first generation interferometers will detect reionization with a 2000h observation.  For MWA128T and PAPER, the peak S/N should be of order unity.  Robust detections are only likely if we can effectively mitigate foregrounds on large ($k<0.1$ Mpc$^{-1}$) scales, or if contiguous frequency coverage allows us to sum detections over several frequency (i.e. redshift) bins.  Stronger detections, with S/N $\gsim10$ are likely with LOFAR.  On the other hand, the continuous bandpass extending to $z\approx18$ allows MWA to detect a broader range of X-ray heating models, compared to the more limited bandpasses of PAPER and LOFAR.  Reasonable models of X-ray heating could be detectable at S/N of unity.  The prospects for detecting both reionization and X-ray heating are much improved for MWA with an extension to 256 tiles.  The SKA will be a huge improvement over first generation instruments.

\vskip+0.5in

We thank A. Liu for providing the 128 element PAPER specifications, HERA specifications, as well as for comments on a draft version of this work.  We also thank A. Loeb for constructive comments on the draft, as well as J. Dillon and L. Koopmans for fruitful conversations. We thank F. Pacucci for assistance with matplotlib.  A. E.-W. and J. H. thank the Massachusetts Institute of Technology School of Science for support. This material is based upon work partially supported by the National Science Foundation Graduate Research Fellowship (to A. E-W.) under Grant No. 1122374.

\bibliographystyle{mn2e}
\bibliography{ms}

\end{document}

%% file: defns.tex
\newcommand{\cmfast}{\textsc{\small 21CMFAST}}

\newcommand{\Ts}{T_{\rm S}}

\newcommand{\nf}{x_{\rm HI}}
\newcommand{\avenf}{\bar{x}_{\rm HI}}

\newcommand{\lya}{Ly$\alpha$}

\newcommand{\Msun}{M_\odot}
\newcommand{\Tvir}{T_{\rm vir}}

\newcommand{\Tcmb}{T_\gamma}
\newcommand{\delT}{\delta T_b}

\newcommand{\delNL}{\delta_{\rm nl}}
\newcommand{\Mmin}{M_{\rm min}}
\newcommand{\Mwdm}{m_{\rm wdm}}

\newcommand\lsim{\mathrel{\rlap{\lower4pt\hbox{\hskip1pt$\sim$}}
        \raise1pt\hbox{$<$}}}
\newcommand\gsim{\mathrel{\rlap{\lower4pt\hbox{\hskip1pt$\sim$}}
        \raise1pt\hbox{$>$}}}
\def\myputfigure#1#2#3#4#5%
{\vskip#5pt\makebox[0pt]{\hskip#2in
\includegraphics[width=#3\textwidth]{#1}}\vskip#4pt\hfill}

